\documentclass[twocolumn,showpacs,nofootinbib,preprintnumbers,prd]{revtex4-1}

\usepackage{amsmath}
\usepackage{amsfonts}
\usepackage{amssymb}
\usepackage{graphicx}
\usepackage[titletoc]{appendix}
\usepackage{color}
\usepackage{hyperref}
\usepackage{cleveref}
\usepackage[rightcaption]{sidecap}
\usepackage{subcaption}
\usepackage{comment}

\usepackage{mathtools}

\usepackage{dcolumn}

\usepackage{array}
\usepackage{ctable}
\usepackage{multirow}
\usepackage{siunitx}
\usepackage{tabularx}
\usepackage{booktabs}
\usepackage{supertabular}
\usepackage{verbatim}

\usepackage{cancel}
\usepackage{graphicx}
\usepackage{dcolumn}
\usepackage{bm}
\usepackage{braket}

\begin{document}

\title{Consequences of non-minimal coupling for mass mixing in spontaneous baryogenesis}

\author{Mattia Dubbini}
\email{mattia.dubbini@studenti.unicam.it}
\affiliation{Universit\`a di Camerino, Via Madonna delle Carceri, Camerino, 62032, Italy.}

\author{Orlando Luongo}
\email{orlando.luongo@unicam.it}
\affiliation{Universit\`a di Camerino, Via Madonna delle Carceri, Camerino, 62032, Italy.}
\affiliation{Department of Nanoscale Science and Engineering, University at Albany-SUNY, Albany, New York 12222, USA.}
\affiliation{Istituto Nazionale di Astrofisica (INAF), Osservatorio Astronomico di Brera, 20121 Milano, Italy.}
\affiliation{Istituto Nazionale di Fisica Nucleare (INFN), Sezione di Perugia, Perugia, 06123, Italy,}
\affiliation{Al-Farabi Kazakh National University, Al-Farabi av. 71, 050040 Almaty, Kazakhstan.}

\author{Marco Muccino}
\email{marco.muccino@unicam.it}
\affiliation{Universit\`a di Camerino, Via Madonna delle Carceri, Camerino, 62032, Italy.}
\affiliation{Al-Farabi Kazakh National University, Al-Farabi av. 71, 050040 Almaty, Kazakhstan.}
\affiliation{ICRANet, Piazza della Repubblica 10, 65122 Pescara, Italy.}

\date{\today}

\begin{abstract}
We investigate the impact of a non-minimal Yukawa-like coupling between curvature and inflaton field within the \emph{spontaneous baryogenesis} background. We demonstrate that this coupling leads to a significant enhancement in particle production, even for small values of the coupling constant $\xi$. Assuming a perfectly homogeneous and isotropic universe during the reheating phase, we study the inflaton decay into fermion-antifermion pairs by means of a semiclassical approach, treating fermions as quantized fields and considering the inflaton and the Ricci scalar as classical quantities. We adopt the simplest approach in which the inflaton is minimally coupled to baryons, and non-minimally with gravity. In particular, we solve the equations of motion for the inflaton to first order in perturbation theory, with $\xi$ serving as perturbative parameter. Afterwards, we compute the difference in the number densities of baryons and antibaryons produced through the inflaton decay into fermion-antifermion pairs. We show that the non-minimal coupling term \emph{de facto} increases inflaton mass, letting fermion-antifermion decays be more probable, and thus enhancing the overall baryogenesis process. As a further outcome, we find that the non-minimal Yukawa coupling also leads to a renormalization of the inflaton mass and weakly influences the bounds over the gravitational constant. Finally, since the fermionic fields appear not to be mass eigenstates, we specialize the mass-mixing between them only. To this end, we thus include the effects of mass-mixing and cosmic expansion into our calculations. Physical consequences of baryon production are therefore explored.
\end{abstract}

\maketitle
\tableofcontents

\section{Introduction}\label{sec:level1}

Baryogenesis is assumed to generate the baryon asymmetry in the very early universe \cite{Cline:2006ts,Dimopoulos:1978pw,Cohen:1990it,Farrar:1993hn}. Across the decades, various attempts have been formulated to explain the observed matter–antimatter asymmetry, namely the baryon predominance over antibaryons \cite{Riotto:1998bt,Bodeker:2020ghk}.

Among the various baryogenesis scenarios, leptogenesis currently provides one of the most compelling and widely supported frameworks \cite{Riotto:1998bt,Fong:2012buy,Davidson:2008bu,Bodeker:2020ghk}. There, a lepton asymmetry is first generated in the early universe through the CP-violating decays of heavy Majorana neutrinos. The model adopts the seesaw mechanism for neutrino mass generation \cite{Dey:2021ecr,Qi:2024pqe,Okada:2012fs,Garny:2006xn} and, consequently, induces -- via lepton asymmetry -- a baryon asymmetry through non-perturbative electroweak processes\footnote{Those are known as sphalerons, violating baryon plus lepton number, $B+L$, while conserving $B-L$.}\cite{Morrissey:2012db}.

Albeit leptogenesis may provide a natural link between the observed matter–antimatter asymmetry and the origin of neutrino masses \cite{Buchmuller:2003gz}, baryogenesis can be constructed in several other ways. Promising efforts consist, e.g. in the widely-consolidate \emph{spontaneous baryogenesis} \cite{Dolgov:1994zq,Dolgov:1996qq,Luongo:2021gho} and \emph{gravitational baryogenesis} \cite{Davoudiasl:2004gf,Arbuzova:2023rri,Sadjadi:2007dx}.

Spontaneous baryogenesis is rooted in the spontaneous breaking of a global $U(1)$ symmetry, that  initially guarantees the total baryon number conservation \cite{Balazs:2014eba}. Accordingly, breaking the symmetry implies the baryon number conservation to be violated, furnishing a baryon asymmetry typically produced during the reheating phase, as the pseudo Nambu–Goldstone boson associated with the broken $U(1)$ symmetry -- identified with the inflaton -- undergoes baryon number-violating decays. The model postulates that the inflaton decays into fermion–antifermion pairs via interactions that explicitly violate baryon number \cite{Racker:2014yfa}.

Conversely, the gravitational baryogenesis is a more recent model that predicts the baryon asymmetry production as due to the coupling between the derivative of the Ricci scalar and a generic matter current that can be either fermionic or scalar. Baryon asymmetry production easily comes out from it, even in non-minimal coupling regime  \cite{Goodarzi:2023ltp}. However, it shows some problems concerning the stability of the Ricci scalar, whose solutions in standard gravity may be divergent\footnote{To this end, involving extended theories of gravity can be helpful to make the solutions stable, see e.g. Ref.~\cite{Arbuzova:2023rri}.} \cite{Arbuzova:2017zby,Mojahed:2024mvb}.

Motivated by the above considerations, we here investigate the spontaneous baryogenesis in a scenario in which the couplings with strong gravity regime cannot be neglected. More precisely, there is consensus in exploring non-minimal inflationary scenarios, which remain viable in describing inflation according to Planck satellite measurements \cite{Planck:2018jri}.\footnote{It has been argued that non-minimal couplings can be associated with possible solutions to the cosmological constant problem \cite{Belfiglio:2022qai}, by extending quasiquintessence scenarios \cite{Belfiglio:2023rxb,Luongo:2024opv,Belfiglio:2024swy}, where scalar field Lagrangian acts as matter-like with pressure \cite{Luongo:2018lgy,DAgostino:2022fcx}.}. In particular, we investigate the presence of a non-minimal Yukawa-like coupling between inflaton and Ricci curvature, both considered as classical fields. There, the inflaton mass may be increased by non-minimal couplings, enhancing the baryon asymmetry as consequence of more probable inflaton decay into fermion-antifermion pairs. We thus explore how this coupling affects the baryon asymmetry and its overall production. To do so, we solve the inflaton equations of motion at first order of perturbation theory, ensuring the non-minimal coupling is small. Accordingly, the first-order correction to the produced baryon asymmetry is obtained. Moreover, a first-order correction to the current model of spontaneous baryogenesis, induced by the non-minimal coupling, is also identified. In addition, as second byproduct, we show that the presence of the non-minimal coupling between gravity and inflaton renormalizes the inflaton mass itself, opening new avenues toward the inflaton nature. Afterwards, the mass mixing is then computed. In this respect, since the Ricci curvature is a classical field, it acts to rescale inflaton's mass, as above stated. The latter is a classical field as well, and thus is not considered to contribute to the mass terms involved in baryon creation. As a consequence, the mass mixing calculation is analogous to the background, when the non-minimal term vanishes. Consequences and limitations on the observable asymmetry are therefore physically discussed. In this respect, we demonstrate that the non-minimal coupling can either reduce or enhance the mixing, depending on the sign of the coupling constant, $\xi$. In contrast to the minimal case, the non-minimal effects therefore appear generally non-negligible and might be incorporated into the spontenous baryogenesis scenario. In view of our energy scales, in fact, we show that there is no reason \emph{a priori} to discard non-minimal couplings that, instead, appear practically unavoidable.

The paper is outlined as follows. In Sec.~\ref{sec:level2}, the spontaneous baryogenesis is reviewed, discussing the corresponding broken phase Lagrangian and its main consequences. In Sec.~\ref{sec:level3}, the full Lagrangian including the non-minimal coupling is reported. This model reduces to former zero order paradigm by setting $\xi=0$.
The equations of motion for inflaton, fermions and gravitational field are found. In Sec.~\ref{sec:level4}, the equations of motion for the inflaton are solved, at first order perturbation, with $\xi$ being the perturbative parameter. Afterwards, in Sec.~\ref{sec:level5} the baryon asymmetry is computed. Finally, in Sec.~\ref{sec:level6}, the mass-mixing and the expansion of the universe are discussed, while in Sec.~\ref{sec:level7}, we discuss conclusions and perspectives of our approach.

\section{Spontaneous baryogenesis}\label{sec:level2}

Baryogenesis is commonly ensured by satisfied the Sakharov conditions \cite{Sakharov:1967dj}, i.e.,
\begin{itemize}
    \item[-] violation of the baryon number $B$, to justify the observed baryon excess;
    \item[-] violation of the discrete symmetries $C$ and $CP$, as necessary consequence to ensure that baryon number-violating processes work differently for matter and antimatter \cite{Dimopoulos:1978kv};
    \item[-] departures from the thermal equilibrium \cite{Weinberg:1979bt}.
\end{itemize}

In the spontaneous baryogenesis, the baryon asymmetry is produced as a consequence of a $U(1)$-symmetry breaking in the early universe. The pseudo Nambu-Goldstone boson produced by this symmetry breaking decays into fermion-antifermion pairs and generate the baryon asymmetry \cite{DeSimone:2016ofp}. Spontaneous baryogenesis can imply dark matter production \cite{Racker:2014yfa,Luongo:2021gho} and effects of mass mixing, leading to observable signature in the primordial stages of universe's evolution, as it will be clarified later.

In this scenario, the boson acts as inflaton, whereas the broken phase Lagrangian reads\footnote{We use a metric signature $(+,-,-,-)$.}
\begin{equation}
\begin{split}
    \mathcal{L}&=\frac{f^2}{2}(\partial_{\mu}\theta)(\partial^{\mu}\theta)-V(\theta)+\overline{Q}(i\gamma^{\mu}\partial_{\mu}-m_Q)Q+\\&+\overline{L}(i\gamma^{\mu}\partial_{\mu}-m_L)L+\frac{gf}{\sqrt{2}}(\overline{Q}Le^{i\theta}+\overline{L}Qe^{-i\theta}),
\end{split}
\label{eq1}
\end{equation}
where $f$ is the energy scale at which the global symmetry is spontaneously broken, $g$ is the coupling constant of the interaction between the fermions, $\theta$ is the inflaton, and $Q$ and $L$ are the fermionic fields, with the first one carrying the baryonic number. Moreover, $V(\theta)=\Lambda^4[1-\cos(\theta)]$ is the \emph{natural inflation} potential \cite{Stein:2021uge} with $f/M_{\text{Pl}}=\mathcal{O}(1)$ and $f/\Lambda\sim 10^3-10^6$\cite{Adams:1992bn}. In the following, we arbitrarily single out $f/\Lambda\sim 10^3$. This choice is not relevant in describing the overall process. Indeed, another ratio $f/\Lambda$ would simply change the values of the coupling constants $\xi$ and $g$ for which perturbative regime is valid, being compatible with the experimental baryon asymmetry.

The choice of $V(\theta)$ appears convenient as it resembles a quadratic term in $\theta$, compatible with observations while coupled non-minimally with $R$ \cite{Opferkuch:2019zbd}. For small oscillations of the inflaton, the potential $V(\theta)$ is expanded around its minimum $\theta=0$ up to the second order, so to have $V(\theta)\simeq (1/2)m^2f^2\theta^2$, where $m=\Lambda^2/f$ is the bare inflaton mass. From the constraints, previously set on $f$ and $\Lambda$, we obtain $m\sim 10^{-6}f$.

The last term in Eq.~\eqref{eq1} is responsible for the baryon asymmetry production. In order to compute it, we first have to solve the equation of motion of the inflaton. In this respect, it is convenient to rewrite the Lagrangian in Eq.~\eqref{eq1} by the rotation, $Q\,e^{i\theta}$, leading to
\begin{equation}
\begin{split}
    \mathcal{L}&=\frac{f^2}{2}(\partial_{\mu}\theta)(\partial^{\mu}\theta)-V(\theta)+\overline{Q}(i\gamma^{\mu}\partial_{\mu}-m_Q)Q+\\&+\overline{L}(i\gamma^{\mu}\partial_{\mu}-m_L)L+\frac{gf}{\sqrt{2}}(\overline{Q}L+\overline{L}Q)-(\partial_{\mu}\theta) J^{\mu},
\end{split}
\label{eq4}
\end{equation}
where $J^{\mu}=\overline{Q}\gamma^{\mu}Q$ is the baryonic current deriving from the $U(1)$-symmetry, that is not conserved in this phase. Starting from the Lagrangian in Eq.~\eqref{eq4} and defining the decay rate of the inflaton $\Gamma=g^2\Omega/(8\pi)$, one can derive the equation of motion for the inflaton\footnote{The term $3H\dot{\theta}$ is not directly derived from Eq. (\ref{eq2}), but it is added to take into account the expansion of the universe.}:
\begin{equation}
    \ddot{\theta}+(3H+\Gamma)\dot{\theta}+\Omega^2\theta=0.
\label{eq5}
\end{equation}
In the equation above, the renormalized mass $\Omega$ is defined with respect to the bare mass $m$ as
\begin{equation}\label{eq6}
m^2=\Omega^2\bigg{[}1+\frac{g^2}{4\pi}\lim_{\omega\to+\infty}\ln\bigg{(}\frac{2\omega}{\Omega}\bigg{)}\bigg{]}.
\end{equation}
The inflaton field is assumed to be weakly coupled to fermions, thus the coupling constant is $g\ll1$ and consequently $\Gamma\ll\Omega$. In his limit and neglecting the Hubble friction, Eq.~\eqref{eq5} admits the following simple solution
\begin{equation}
    \theta(t)=\theta_Ie^{-\Gamma t/2}\cos(\Omega t),
\label{eq7}
\end{equation}
in the limit $\Gamma\ll\Omega$. This expression for $\theta$ is then used to compute the baryon asymmetry.

The average number density $n$ of particles-antiparticles pairs produced by the decay of a homogeneous classical scalar field is given by
\begin{equation}
    n=\frac{1}{V}\sum_{s1,s2}\int\frac{d^3p_1}{(2\pi)^32p_1^0}\frac{d^3p_2}{(2\pi)^32p_2^0}|\mathcal A|^2,
\label{eq2}
\end{equation}
where $\mathcal A$ is the single pair production amplitude.

If we set $n(Q,\overline{L})$ as the number density of produced baryons, $n_b$, and $n(\overline{Q},L)$ as the number density of produced antibaryons, $n_{\overline{b}}$, then it is possible to find the baryon asymmetry produced $n_B=n_b-n_{\overline{b}}$. As explicitly shown in Refs.~\cite{Dolgov:1994zq,Dolgov:1996qq}, if one neglects both the expansion of the universe and the mass-mixing, the net produced baryon asymmetry turns out to give
\begin{equation}
    n_B=\frac{1}{16\pi}\Omega g^2f^2\theta_I^3,
\label{eq3}
\end{equation}
where $\theta_I$ is the value of the inflaton at the beginning of reheating. If we do not consider the expansion of the universe, we have  $\theta_I=\sqrt{3/(4\pi)}\Gamma M_{\text{Pl}}/(f\Omega)$. Conversely, considering the universe expansion provides a correction by a factor $8\pi/g^2$ and, moreover, the mass-mixing is shifted by a factor $[(1-\epsilon^2)/(1+\epsilon^2)]^2$, where $\epsilon=\sqrt{2}gf/[\Delta m+\sqrt{\Delta m^2+2g^2f^2}]$, with $\Delta m=m_Q-m_L$.

Below, we demonstrate that non-minimally correcting the Lagrangian leads to modifying the baryon asymmetry in Eq.~\eqref{eq3} roughly of a factor $\sim 1-2\pi\Xi/(3g^2)$, with $|\Xi|\propto\xi$, excluding the cosmic expansion at first glance. However,  restoring it at the end of our work implies that our final finding results to be even larger, showing that the expansion enhances baryon asymmetry, in line with current outcomes related to particle production at primordial times, see e.g. Refs.~\cite{Belfiglio:2025chv,Belfiglio:2024swy,Belfiglio:2024xqt,Belfiglio:2023moe,Belfiglio:2023moe,Belfiglio:2023rxb,Belfiglio:2022qai,Belfiglio:2022yvs,Belfiglio:2022cnd,Aviles:2011sfa}.

\section{Non-minimally coupled baryogenesis}\label{sec:level3}

In this section, we rewrite Eq.~\eqref{eq4} by adding a non-minimal term that couples geometry and scalar fields. There is no reason to exclude \emph{a priori} such a term, as due to the high energy scales at which baryogenesis occurs.

In so doing, we pursue the procedure previously discussed, finding out inflationary equations of motion that appear different from Eq.~\eqref{eq5}, as due to the presence of a non-linear term proportional to $\xi$, thst can be solved adopting a perturbative approach at first order.

Let $\mathcal{L}_{\text{m}}$ be the (zero order) Lagrangian in Eq.~\eqref{eq4}. Such a Lagrangian cannot be used for our purposes, because it is a Lorentz scalar only for flat spacetime.
In fact, when introducing in the theory a non-minimal coupling between gravity and inflaton, a time-dependent expression for the Ricci scalar appears in the equation of motion for the inflaton. Thus, we have to work in a curved spacetime, that in our case is the Friedmann-Lemaître-Robertson-Walker (FLRW) spacetime.
The Lagrangian $\mathcal{L}_{\text{m}}$ generalized for the FLRW spacetime is
\begin{equation}
\begin{split}
    \mathcal{L}_{\text{m}}&=\frac{f^2}{2}(\nabla_{\mu}\theta)(\nabla^{\mu}\theta)-V(\theta)+\\&+\frac{i}{2}\bigg{(}\overline{Q}\gamma^{\alpha}\nabla_{\alpha}Q-\nabla_{\alpha}(\overline{Q}\gamma^{\alpha})Q\bigg{)}-m_Q\overline{Q}Q+\\&+\frac{i}{2}\bigg{(}\overline{L}\gamma^{\alpha}\nabla_{\alpha}L-\nabla_{\alpha}(\overline{L}\gamma^{\alpha})L\bigg{)}-m_L\overline{L}L+\\&+\frac{gf}{\sqrt{2}}(\overline{Q}L+\overline{L}Q)-(\nabla_{\mu}\theta) J^{\mu},
\end{split}
\label{eq8}
\end{equation}
where the covariant derivative for fermions and the Dirac matrices in the FLRW spacetime are defined through the tetrad formalism. Because we perfectly know how to write the fermion Lagrangian in the flat Minkowski spacetime with metric tensor $\eta_{ab}$, we consider tetrads such that $g_{\mu\nu}=\eta_{ab}e^a_{\mu}e^b_{\nu}$ and $\eta_{ab}=g_{\mu\nu}e^{\mu}_ae^{\nu}_b$, where the metric tensor for a spatially-flat FLRW universe in cartesian coordinates is $g_{\mu\nu}=dt^2-a^2(t)(dx^2+dy^2+dz^2)$.

Adopting the tetrad representation with
\begin{equation}
    e^a_{\mu}=
    \begin{pmatrix}
    1 & 0 & 0 & 0 \\
    0 & a & 0 & 0 \\
    0 & 0 & a & 0 \\
    0 & 0 & 0 & a
    \end{pmatrix},
\label{eq10}
\end{equation}
and its inverse matrix $e^{\mu}_a$, the Dirac matrices in a FLRW spacetime are defined as $\gamma^{\mu}=e^{\mu}_a\gamma^a$, with Dirac matrices $\gamma^a$ in the flat Minkowski spacetime.
Moreover, we can define the covariant derivative for a spinor as
\begin{equation}
\nabla_{\mu}\psi=\partial_{\mu}\psi+\chi_{\mu}\psi=\partial_{\mu}\psi+\frac{1}{8}\omega_{\mu bc}[\gamma^b,\gamma^c]\psi,
\label{eq11}
\end{equation}
with the spin connection $\omega_{\mu bc}$ defined as
\begin{equation}
\omega_{\mu a b}=\eta_{ac}e^c_{\nu}e^{\sigma}_b\Gamma^{\nu}_{\sigma\mu}+\eta_{ac}e^c_{\nu}\partial_{\mu}e^{\nu}_b.
\label{eq12}
\end{equation}
It is not difficult to prove that $\chi_0=0$ and for each $i=1,2,3$, we have $\chi_i=\dot{a}\gamma_i\gamma_0/2$.
Bearing this in mind, we end up with our \emph{complete Lagrangian description}
\begin{equation}
    \mathcal{L}=\mathcal{L}_{\text{m}}-\frac{M^2_{\text{Pl}}}{16\pi}R-\frac{f^2}{2}\xi\theta^2R.
\label{eq13}
\end{equation}
containing the additional non-minimal coupling $\sim \theta^2R$.

\subsection{\label{subsec}Equations of motion}

The equations of motion for the inflaton $\theta$ and the fermions $Q$ and $L$ are, respectively,
\begin{subequations}
    \begin{align}
    &f^2\nabla^2\theta+V'(\theta)+f^2\xi R \theta=\braket{\nabla_{\mu}J^{\mu}}, \label{eq14}\\
    &i\gamma^{\mu}\nabla_{\mu}Q-m_QQ-(\nabla_{\mu}\theta)\gamma^{\mu}Q=-\frac{gf}{\sqrt{2}}L,\label{eq16}\\
    &i\gamma^{\mu}\nabla_{\mu}L-m_LL=-\frac{gf}{\sqrt{2}}Q,\label{eq17}
    \end{align}
\end{subequations}
becoming in a FLRW universe\footnote{In principle, as done for the field $\theta$, in a FLRW universe we take fields only depending on time. Consequently, the gradient term in the equations of motion is negligible. However, we need to quantize the fermionic fields. In canonical quantization, fields are functions of a point in the spacetime, included the spatial part, therefore the gradient term cannot be neglected anymore.}, respectively,
\begin{subequations}
    \begin{align}
    &f^2(\ddot{\theta}+3H\dot{\theta})+f^2\xi R\theta+V'(\theta)=\braket{\nabla_{\mu}J^{\mu}},\label{eq15}\\
    &i\gamma^0\!\left(\dot{Q}+\frac{3H}{2}Q\right)\!+\frac{i}{a}\gamma^i\partial_iQ-m_QQ-\dot{\theta}\gamma^0Q=-\frac{gf}{\sqrt{2}}L,\label{eq18}\\
    &i\gamma^0\!\left(\dot{L}+\frac{3H}{2}L\right)\!+\frac{i}{a}\gamma^i\partial_iL-m_LL=-\frac{gf}{\sqrt{2}}Q.
\label{eq19}
    \end{align}
\end{subequations}
To compute the inflaton equations of motion in Eq.~\eqref{eq15}, the \emph{vacuum expectation value} (VEV) $\braket{\nabla_{\mu}J^{\mu}}$ might be evaluated and, so, we remarkably notice that
\begin{equation}
\nabla_{\mu}J^{\mu}=\frac{igf}{\sqrt{2}}(\overline{Q}L-\overline{L}Q).
\label{eq20}
\end{equation}

Pursuing the same procedure of the background model, we first solve the equations of motion for fermions by using perturbation theory over the small parameter $g$. In doing this, here we assume that the universe expansion is negligible, thus $a(t)$ varies slowly over cosmic time.
Hence, let $Q_0(t,\vec{x})$ and $L_0(t,\vec{x})$ be the zero-order solutions of Eqs.~\eqref{eq16}--\eqref{eq17}, respectively\footnote{The calculation of the solution of the free Dirac equation in FLRW spacetime can be found in Appendix \ref{app0}.},
\begin{subequations}
\begin{align}
\nonumber
&Q_0(t,\vec{x})=a^{-\frac{3}{2}}(t)e^{-i\theta(t)}\sum_s\int\frac{d^3p}{(2\pi)^3}\frac{1}{\sqrt{2E_p}}\times \\
\label{eq21}
&\!\left[u(\vec{p},\!s)\hat a(\vec{p},\!s)e^{-i(\!E_pt-\vec{p}\cdot\vec{x})}\!+\!v(\vec{p},\!s)\hat b^{\dag}\!(\vec{p},\!s)e^{i(\!E_pt-\vec{p}\cdot\vec{x})}\!\right]\!\\
\nonumber
&L_0(t,\vec{x})=a^{-\frac{3}{2}}(t)\sum_s\int\frac{d^3p}{(2\pi)^3}\frac{1}{\sqrt{2E_p}}\times \\
&\!\left[u(\vec{p},\!s)\hat c(\vec{p},\!s)e^{-i(\!E_pt-\vec{p}\cdot\vec{x})}\!+\!v(\vec{p},\!s)\hat d^{\dag}\!(\vec{p},\!s)e^{i(\!E_pt-\vec{p}\cdot\vec{x})}\!\right]\!\label{eq22}
\end{align}
\end{subequations}
where $\vec{p}$ is the three-momentum, $E_p$ is the energy, and $u$ and $v$ are spinors. Further, we indicate with $\hat a$ and $\hat c$ the annihilation operators, whereas with $\hat b^\dagger$ and $\hat d^\dagger$ the creation operators.

The first order solutions $Q_1(t,\vec{x})$ and $L_1(t,\vec{x})$ of Eqs.~\eqref{eq16}--\eqref{eq17} are, respectively,
\begin{subequations}
\begin{align}
\nonumber
&Q_1(t,\vec{x})=Q_0(t,\vec{x})-a^{-\frac{3}{2}}(t)e^{-i\theta(t)}\frac{gf}{\sqrt{2}}\times \\
&\times \int dt'\int d^3y G_Q(t,\vec{x};t',\vec{y})L_0(t',\vec{y})a^{\frac{3}{2}}(t')e^{i\theta(t')}, \label{eq23}\\
\nonumber
&L_1(t,\vec{x})=L_0(t,\vec{x})-a^{-\frac{3}{2}}(t)\frac{gf}{\sqrt{2}}\times \\& \times \int dt'\int d^3y G_L(t,\vec{x};t',\vec{y})Q_0(t',\vec{y})a^{\frac{3}{2}}(t').
\label{eq24}
\end{align}
\end{subequations}

To have a non-zero VEV $\braket{\nabla_{\mu}J^{\mu}}$, first-order solutions in Eqs.~\eqref{eq23}--\eqref{eq24} have to be used leading to
\begin{equation}
\braket{\nabla_{\mu}J^{\mu}}= \frac{g^2f^2}{4\pi a^{3}(t)}\bigg{[}-\frac{\Omega\dot{\theta}}{2}+\Omega^2\theta\lim_{\omega\to+\infty}\ln\bigg{(}\frac{2\omega}{\Omega}\bigg{)}\bigg{]}.
\label{eq25}
\end{equation}
We recover the background model, as expected, since the calculation only involves fermions, whose Lagrangian is kept unchanged, with the time dependence $a^{-3}(t)$, as expected for a FLRW spacetime.

For the gravitational sector, we immediately find
\begin{equation}
\begin{split}
    &\frac{M^2_{\text{Pl}}}{8\pi}G_{\mu\nu}+f^2\xi\bigg{(}G_{\mu\nu}+g_{\mu\nu}\nabla^2-\nabla_{\mu}\nabla_{\nu}\bigg{)}\theta^2=T_{\mu\nu}
\end{split}
\label{eq28}
\end{equation}
where $G_{\mu\nu} = R_{\mu\nu}-(1/2)g_{\mu\nu}R$ is the Einstein tensor and $T_{\mu\nu}$ is the energy-momentum tensor associated to $\mathcal{L}_{\text{m}}$
\begin{align}
\nonumber
&T_{\mu\nu}=-\frac{f^2}{2}g_{\mu\nu}(\nabla^{\alpha}\theta)(\nabla_{\alpha}\theta)+g_{\mu\nu}V(\theta)+f^2(\nabla_{\mu}\theta)(\nabla_{\nu}\theta)+\\
\nonumber
&-\frac{i}{2}g_{\mu\nu}\bigg{(}\overline{Q}\gamma^{\alpha}\nabla_{\alpha}Q-\nabla_{\alpha}(\overline{Q}\gamma^{\alpha})Q\bigg{)}+g_{\mu\nu}m_Q\overline{Q}Q+\\
\nonumber
&+\frac{i}{4}\bigg{(}\overline{Q}\gamma_{\mu}\nabla_{\nu}Q+\overline{Q}\gamma_{\nu}\nabla_{\mu}Q-\nabla_{\nu}(\overline{Q}\gamma_{\mu})Q-\nabla_{\mu}(\overline{Q}\gamma_{\nu})Q\bigg{)}+\\
\nonumber
&-\frac{i}{2}g_{\mu\nu}\bigg{(}\overline{L}\gamma^{\alpha}\nabla_{\alpha}L-\nabla_{\alpha}(\overline{L}\gamma^{\alpha})L\bigg{)}+g_{\mu\nu}m_L\overline{L}L+\\
\nonumber
&+\frac{i}{4}\bigg{(}\overline{L}\gamma_{\mu}\nabla_{\nu}L+\overline{L}\gamma_{\nu}\nabla_{\mu}L-\nabla_{\nu}(\overline{L}\gamma_{\mu})L-\nabla_{\mu}(\overline{L}\gamma_{\nu})L\bigg{)}+\\
\nonumber
&+g_{\mu\nu}(\nabla_{\alpha}\theta)J^{\alpha}-(\nabla_{\mu}\theta)J_{\nu}+\\
&-(\nabla_{\nu}\theta)J_{\mu}-g_{\mu\nu}\frac{gf}{\sqrt{2}}(\overline{Q}L+\overline{L}Q).
\label{eq29}
\end{align}
By taking the trace of Eq.~\eqref{eq28}, we get
\begin{equation}
    -\frac{M^2_{\text{Pl}}}{8\pi}R-f^2\xi(R-3\nabla^2)\theta^2=T^{\mu}_{\mu},
\label{eq30}
\end{equation}
where $T^{\mu}_{\mu}$ is the trace of Eq.~\eqref{eq29}, that is
\begin{equation}
\begin{split}
&T^{\mu}_{\mu}=-f^2(\nabla^{\alpha}\theta)(\nabla_{\alpha}\theta)+4V(\theta)+\\&-\frac{3i}{2}\bigg{(}\overline{Q}\gamma^{\mu}\nabla_{\mu}Q-\nabla_{\mu}(\overline{Q}\gamma^{\mu})Q\bigg{)}+4m_Q\overline{Q}Q+\\&-\frac{3i}{2}\bigg{(}\overline{L}\gamma^{\mu}\nabla_{\mu}L-\nabla_{\mu}(\overline{L}\gamma^{\mu})L\bigg{)}+4m_L\overline{L}L+\\&+2(\nabla_{\alpha}\theta)J^{\alpha}-4\frac{gf}{\sqrt{2}}(\overline{Q}L+\overline{L}Q).
\end{split}
\label{eq31}
\end{equation}
Such a trace can be simplified by using the equations of motion for fermions. It becomes
\begin{equation}
\begin{split}
&T^{\mu}_{\mu}=-f^2(\nabla^{\alpha}\theta)(\nabla_{\alpha}\theta)+4V(\theta)+m_Q\overline{Q}Q+\\
& m_L\overline{L}L+-(\nabla_{\alpha}\theta)J^{\alpha}-\frac{gf}{\sqrt{2}}(\overline{Q}L+\overline{L}Q).
\end{split}
\label{eq32}
\end{equation}
Finally, Eq.~\eqref{eq30} can be written as
\begin{equation}
\begin{split}
&-\frac{M^2_{\text{Pl}}}{8\pi}R-f^2\xi(R-3\nabla^2)\theta^2=-f^2(\nabla\theta)^2+4V(\theta)\\
&-(\nabla_{\mu}\theta)J^{\mu}+m_Q\overline{Q}Q+m_L\overline{L}L-\frac{gf}{\sqrt{2}}(\overline{Q}L+\overline{L}Q).
\end{split}
\label{eq33}
\end{equation}

Since the coupling constant is very small, namely $g\ll1$, we expect the fermions produced by the decays of the inflaton in fermion-antifermion pairs to prompt very small masses, roughly of the order of $gf$. Thus, we may handle the approximation of massless fermions. However, it can be shown \cite{Dolgov:1996qq} that non-zero values of $m_Q$ and $m_L$ do not change the solution for $\theta$ significantly,  as long as $m_Q,m_L\ll\Omega$, as we assume below. In this regime, in a spatially flat FLRW spacetime, Eq.~\eqref{eq33} becomes
\begin{equation}
\begin{split}
&R\bigg{(}\frac{M^2_{\text{Pl}}}{8\pi}+f^2\xi\theta^2\bigg{)}-6f^2\xi(\dot{\theta}^2+\theta\ddot{\theta}+3H\theta\dot{\theta})=\\&=f^2\dot{\theta}^2-4V(\theta)+\dot{\theta}\braket{J^0}+\frac{gf}{\sqrt{2}}\braket{\overline{Q}L+\overline{L}Q}.
\end{split}
\label{eq34}
\end{equation}

In computing the VEVs in Eq. (\ref{eq34}), we use once again perturbation theory. In particular, for $\braket{\overline{Q}L+\overline{L}Q}$, we use perturbation theory at first order, having \footnote{The complete set of calculations is reported in Appendix \ref{app1a}.}
\begin{equation}
\frac{gf}{\sqrt{2}}\braket{\overline{Q}L+\overline{L}Q}=a^{-3}(t)\frac{g^2f^2\Omega^2}{8\pi}\theta(t).
\label{eq26}
\end{equation}

Conversely, to compute the VEV $\braket{J^0}$, the zero-order in perturbation theory is enough and, accordingly\footnote{Again, complete calculations are reported in Appendix \ref{app1b}.}
\begin{equation}
\braket{J^0}=\frac{P^3_{\text{max}}}{3\pi^2}a^{-3}(t),
\label{eq27}
\end{equation}
where it appears clear the need of a further cutoff scale, identified by $P_{\text{max}}$, reported within the integral as due to the natural divergence arising from computing energy density in quantum field theories. Hereafter, $P_{\text{max}}$ turns out to be a free parameter of the theory itself. Roughly, we do expect it to lie on the energy scale $f$.

By substituting in Eq.~\eqref{eq34} the VEVs of Eqs.~\eqref{eq26}--\eqref{eq27}, we end up with
\begin{equation}
\begin{split}
&R\bigg{(}\frac{M^2_{\text{Pl}}}{8\pi}+f^2\xi\theta^2\bigg{)}-6f^2\xi(\dot{\theta}^2+\theta\ddot{\theta}+3H\theta\dot{\theta})=\\&=f^2\dot{\theta}^2-4V(\theta)+\dot{\theta}\frac{P^3_{\text{max}}}{3\pi^2}a^{-3}+a^{-3}\frac{g^2f^2\Omega^2}{8\pi}\theta.
\end{split}
\label{eq35}
\end{equation}

For now, we may assume that the expansion of the universe turns out to be negligible with respect to the inflaton decay rate, namely $\Gamma\gg H$. Further, by virtue of the definition of $\theta_I$ given in Sec.~\ref{sec:level2}, we assume $\theta$ of the order of $\Gamma/\Omega\sim g^2$. In agreement with the solution for the inflaton for $\xi=0$, given by Eq.~\eqref{eq5}, we have that $\dot{\theta}\sim\Omega\theta$. This last consideration, together with $\Gamma\gg H$ and $\Omega\gg\Gamma$, yield $\dot{\theta}\sim\Omega\theta\gg\Gamma \theta\gg H\theta$. This means that we are allowed to consider $\dot{\theta}\gg H\theta$, namely the time variation of the inflaton is faster than the one due to the universe expansion. Last but not least, we also assume $a\simeq 1$, that  consists in neglecting the effects of cosmic expansion, which will be restored later on to show their net influence.
In this regime, Eq.~\eqref{eq35} reduces to
\begin{equation}
\begin{split}
&R\bigg{(}\frac{M^2_{\text{Pl}}}{8\pi}+f^2\xi\theta^2\bigg{)}-6f^2\xi(\dot{\theta}^2+\theta\ddot{\theta})=\\&f^2\dot{\theta}^2-2f^2\Omega^2\theta^2+\frac{P^3_{\text{max}}}{3\pi^2}\dot{\theta}+f^2\Omega\Gamma\theta.
\end{split}
\label{eq36}
\end{equation}

In deriving Eq.~\eqref{eq36}, we replace the bare mass $m$ with the renormalized one, $\Omega$, as defined in Eq.~\eqref{eq6}. Renormalization is performed directly at the level of the $\theta$-equations of motion, where the VEV, $\braket{\nabla_{\mu}J^{\mu}}$, is substituted according\footnote{This procedure is formally equivalent to the introduction of a counterterm in the Lagrangian that accounts for mass renormalization. As a result, all instances of the bare mass $m$ can be consistently replaced by the renormalized mass $\Omega$ within the theoretical framework.} to Eq.~\eqref{eq25}. The same constraints previously set on the bare mass $m$ are thus now valid for $\Omega$. In particular, we take $\Omega\sim 10^{-6}f$, as it was for $m$.

Eq.~\eqref{eq36} can be further simplified by evaluating the order of magnitude of each term. To do so, we remark once again that we are considering $f/M_{\text{Pl}}=\mathcal{O}(1)$ and $\Omega\sim 10^{-6}f$. We notice that
\begin{itemize}
\item[-] $M^2_{\text{Pl}}/(8\pi)\gg f^2\xi\theta^2$, since $\xi\theta^2\sim\xi g^4\ll1$;
\item[-] $f^2\dot{\theta}^2\gg 6f^2\xi(\dot{\theta}^2+\theta\ddot{\theta})$, since $\dot{\theta}^2$ is of the same order of magnitude of $\theta\ddot{\theta}$, that is $\Omega^2\theta^2$, and the quantity in the right hand side is thus smaller than the one in the left hand side by a factor $\xi\ll1$;
\item[-] in the second line of Eq.~(\ref{eq36}), the term proportional to $P^3_{\text{max}}$ is much greater than the other three, being it of the order of $f^3\Omega\theta\sim 10^6 f^2\Omega^2g^2$ and the others all of the order of $f^2\Omega^2g^4$.
\end{itemize}
In this way, we obtain a simple explicit time-dependent expression of the Ricci scalar,
\begin{equation}
R=\frac{8\pi}{M^2_{\text{Pl}}}\frac{P^3_{\text{max}}}{3\pi^2}\dot{\theta}.
\label{eq37}
\end{equation}
Since terms proportional to $\xi$ are negligible, we work at order zero and, thus, Eq.~\eqref{eq37} is also the explicit form of the Ricci curvature, obtained in the background model. However, at background level, the Ricci curvature does not affect the equation of motion for the inflaton, as the non-minimal coupling is absent, and thus computing it appears unnecessary.

\section{Dynamical consequences within inflationary stages}\label{sec:level4}

We now insert $\braket{\nabla_{\mu}J^{\mu}}$ from Eq.~\eqref{eq25}, with $a=1$ and the time-dependent Ricci scalar from Eq.~\eqref{eq37}, in Eq.~\eqref{eq15}, neglecting in the latter the Hubble friction with respect to the decay rate. Accordingly, we obtain
\begin{equation}
\begin{split}
&\ddot{\theta}+\Gamma\dot{\theta}+\xi\theta \frac{8\pi}{M^2_{\text{Pl}}}\frac{P^3_{\text{max}}}{3\pi^2}\dot{\theta}+\Omega^2\theta=0.
\end{split}
\label{eq38}
\end{equation}

It is straightforward to check that, for $\xi=0$, we recover Eq.~(\ref{eq5}), with the only difference that in Eq.~(\ref{eq38}) the Hubble friction is already neglected with respect to the decay rate $\Gamma$ of the inflaton.  According to cosmological observations, we focus on $\xi\ll1$ only to guarantee that the gravitational constant remains almost unaltered across the entire process of inflation.

For simplicity, we define the energy parameter
\begin{equation}
\Upsilon=\frac{8\pi}{M^2_{\text{Pl}}}\frac{P^3_{\text{max}}}{3\pi^2},
\label{eq40}
\end{equation}
that is comparable to the energy scale $f$ since  $P_{\text{max}}\sim f\sim M_{\text{Pl}}$.
In this way, the equations of motion for $\theta$ yield
\begin{equation}
\begin{split}
&\ddot{\theta}+\Gamma\dot{\theta}+\Upsilon\xi\theta\dot{\theta}+\Omega^2\theta=0,
\end{split}
\label{eq42}
\end{equation}
dubbed the \emph{Li\'enard equation}, where we need to impose the following initial conditions
\begin{subequations}\label{initialconditions}
    \begin{align}
&\theta(0)=\theta_I,\\
&\dot{\theta}(0)=0,
    \end{align}
\end{subequations}
indicating \emph{de facto} that the inflaton magnitude is initially imposed as a large field value, $\theta_I$, whereas having negligible kinematics at initial stages.

\subsection{Solving the equations of motion}

A general exact solution for such an equation cannot be found easily, albeit there is no need to exactly solve it, as due to the fact that the non-linear term, by constraining $\xi$ as we are going to see, represents a correction to the overall motion.
Hence, we can solve Eq.~(\ref{eq42}) adopting perturbation theory.
In particular, we truncate the series at first order in $\xi$, letting $\theta_0(t)$ be the zero-order solution and $\theta_1(t)$ the first-order solution.

A general solution can be therefore written as
\begin{equation}
\theta(t)=\theta_0(t)+\xi\theta_1(t)+\mathcal{O}(\xi^2).
\label{eq43}
\end{equation}
Bearing this form for $\theta(t)$ in mind, Eq. (\ref{eq42}) becomes
\begin{equation}
\ddot{\theta}_0+\Gamma\dot{\theta}_0+\Omega^2\theta_0+\xi \left(\ddot{\theta}_1+\Gamma\dot{\theta}_1+\Omega^2\theta_1+\Upsilon\theta_0\dot{\theta}_0\right)+\mathcal{O}(\xi^2)=0.
\label{eq44}
\end{equation}
Perturbatively, the above expression acquires the form
\begin{equation}
\begin{cases}
\ddot{\theta}_0+\Gamma\dot{\theta}_0+\Omega^2\theta_0=0,\\
\ddot{\theta}_1+\Gamma\dot{\theta}_1+\Omega^2\theta_1=-\Upsilon\theta_0\dot{\theta}_0,
\end{cases}
\label{eq45}
\end{equation}
\label{eq45}
whose initial conditions extend Eqs. \eqref{initialconditions} as
\begin{equation}
\begin{cases}
\theta_0(0)=\theta_I,\text{\ \ \ }\dot{\theta}_0(0)=0,\\
\theta_1(0)=0,\text{\ \ \ }\dot{\theta}_1(0)=0.
\end{cases}
\label{eq46}
\end{equation}
The zero-order solution is exactly given by Eq. (\ref{eq7}), as expected. The first-order solution is found to be
\begin{equation}
\theta_1(t)=\frac{\Upsilon}{3\Omega}\theta_I^2e^{-\Gamma t/2}\sin(\Omega t)\left[1-e^{-\Gamma t/2}\cos(\Omega t)\right].
\label{eq47}
\end{equation}
We baptize the dimensionless parameter
\begin{equation}
\Xi\equiv\xi\Upsilon\theta_I/\Omega,
\end{equation}
which is of the order of $10^6|\xi| g^2$ and represents a perturbative correction to the background as long as $10^6|\xi|g^2\ll 1$. For practical reasons, we will use this condition to display viable plots.
With this in mind, the full solution of Eq.~(\ref{eq42}) at first-order in perturbation theory is
\begin{equation}
\frac{\theta(t)}{\theta_I}\!=\!\frac{\theta_0(t)}{\theta_I}\!+\! \frac{\Xi}{3}e^{-\Gamma t/2} \sin(\Omega t)\!\left[\!1-e^{-\Gamma t/2}\cos(\Omega t)\!\right]\!.
\label{eq48}
\end{equation}

Eq.~\eqref{eq48} holds for $t\ge0$, when the inflaton starts decaying; for $t<0$, we consider $\theta(t)=\theta_I$. We immediately verify that for $\xi=0$ ($\Xi=0$) we recover the zero-order solution of Eq.~(\ref{eq7}).
Moreover, the order of magnitude of $\Xi$ is a direct consequence of our arbitrariness in selecting $f/\Lambda\sim10^3$. Other bounds, i.e., $f/\Lambda\sim \{10^4,10^5,10^6\}$, yield $\Xi\sim\{10^8,10^{10},10^{12}\} \xi g^2$ in lieu of $10^6\xi g^2$. Consequently, the values of $\xi$ and $g$, guaranteeing the perturbative regime, would change accordingly.

\begin{figure*}
\centering
\includegraphics[width=0.89\hsize,clip]{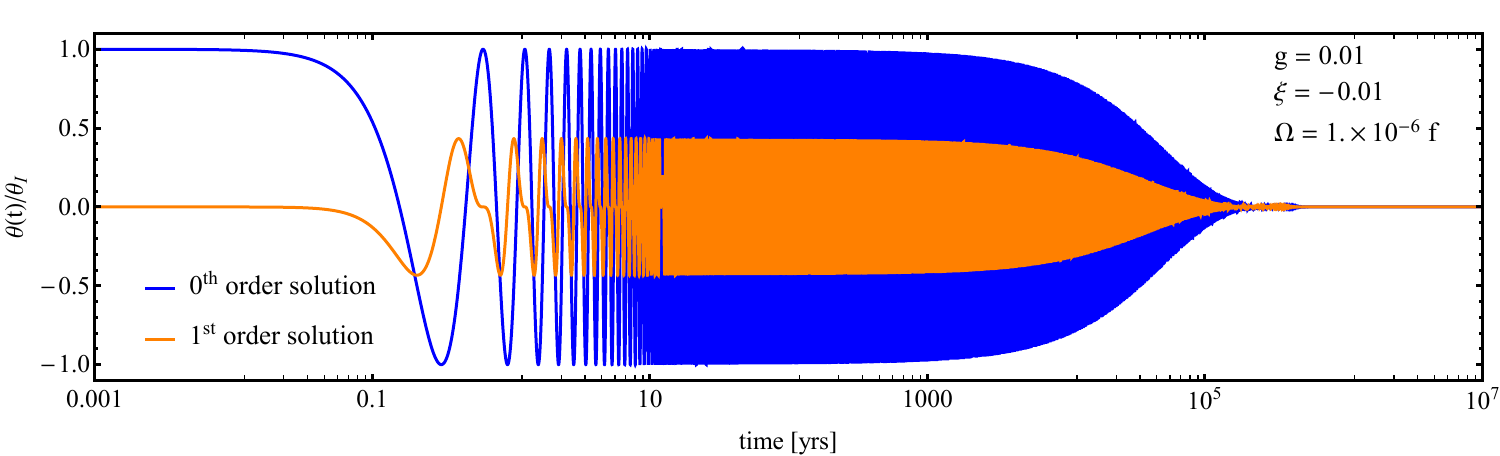}\\
\includegraphics[width=0.89\hsize,clip]{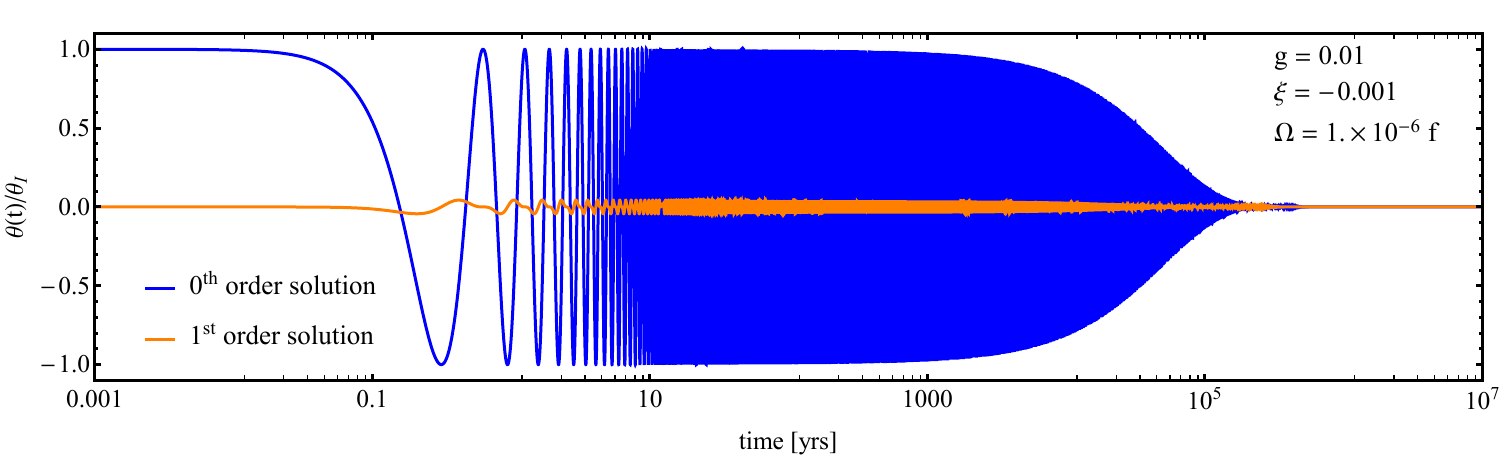}
\caption{Comparison between zero and first order solutions for $\theta(t)$. \emph{Top}: for $g=0.01$, $\Omega=10^{-6}f$, and $\xi=-0.01$, the first order is comparable to the background solution, thus the perturbative method fails. \emph{Bottom}: for $g=0.01$, $\Omega=10^{-6}f$, and $\xi=-0.001$, the first order solution represents a small correction to the background.}
\label{fig1}
\end{figure*}

In Fig.~\ref{fig1} we show the comparison between the first order correction and the zero order solution and check the validity of the perturbative regime. In the top panel of Fig.~\ref{fig1}, we take $g=0.01$ and $|\xi|=0.01$, thus $|\Xi|\sim10^6|\xi|g^2\sim1$. In this case the perturbative method fails, as the first order correction becomes comparable to the background, and Eq.~(\ref{eq48}) is not applicable. Differently, in the bottom panel of Fig.~\ref{fig1}, we take $g=0.01$ and $|\xi|=10^{-3}$, so we get $|\Xi|\sim10^6|\xi|g^2\sim0.1$. In this case, the first order correction represents just a small perturbation to the background, thus, it is licit to work in perturbative regime and Eq.~(\ref{eq48}) is valid.

Nevertheless, in the following we prove that the presence of such a corrective term leads to a relevant correction in terms of particles production. In particular, \emph{the baryon asymmetry produced can be significantly affected by the presence of this first order correction, even though this is small}. In the next section, we thus need to face this point in more detail, showing the way baryogenesis is affected once the non-minimal coupling is included.

\section{Particle production}\label{sec:level5}

We compute the baryon asymmetry $n_B$ as the difference between the number densities of produced baryons ($n_b$) and produced antibaryons ($n_{\overline{b}}$). In particular, we consider $n_b=n(Q,\overline{L})$ and $n_{\overline{b}}=n(L,\overline{Q})$. The average number density of pairs $Q\text{-}\overline{L}$ and $L\text{-}\overline{Q}$ produced by the decay of the inflaton are computed by using Eq.~(\ref{eq2}). At first order in perturbation theory, they are respectively
\begin{subequations}
\begin{align}
\nonumber
&\!n(Q,\overline{L})=\frac{1}{V}\sum_{s_Q,s_L}\int\frac{d^3p_Q}{(2\pi)^32p_Q^0}\frac{d^3p_L}{(2\pi)^32p_L^0} \bigg{|}i\frac{gf}{\sqrt{2}}\times \\
&\!\hat{T}\!\left[\!\int \!d^4x\!\braket{Q(p_Q,\!s_Q\!),\overline{L}(p_L,\!s_L\!|\hat{\overline{Q}}(x)\hat{L}(x)e^{i\theta(x)}\!|0}\!\right]\!\bigg{|}^2\!,\!\label{eq49}\\
\nonumber
&\!n(L,\overline{Q})=\frac{1}{V}\sum_{s_Q,s_L}\int\frac{d^3p_Q}{(2\pi)^32p_Q^0}\frac{d^3p_L}{(2\pi)^32p_L^0}\bigg{|}i\frac{gf}{\sqrt{2}}\times \\
&\!\hat{T}\!\left[\!\int \!d^4x\!\braket{L(p_L,\!s_L\!)\!,\!\overline{Q}(p_Q,\!s_Q\!)|\hat{\overline{L}}(x)\hat{Q}(x)e^{-i\theta(x)\!}|0}\!\right]\!\bigg{|}^2\!,\! \label{eq50}
\end{align}
\end{subequations}
where $\hat T$ is the time-ordering operator.

We perform the full calculations of $n(Q,\overline{L})$ and $n(L,\overline{Q})$ in the ultrarelativistic limit, because, as already discussed, the masses of the fermions produced by the decays of the inflaton are very small in view of the coupling constant $g\ll1$. Using the canonical quantization for fermion fields, we obtain
\begin{subequations}
    \begin{align}
&n(Q,\overline{L})=\frac{g^2f^2}{2\pi^2}\int \omega^2\bigg{|}\int_{-\infty}^{+\infty}dte^{i\theta}e^{2i\omega t}\bigg{|}^2d\omega,\label{eq51}\\
&n(L,\overline{Q})=\frac{g^2f^2}{2\pi^2}\int \omega^2\bigg{|}\int_{-\infty}^{+\infty}dte^{-i\theta}e^{2i\omega t}\bigg{|}^2d\omega.
\label{eq52}
\end{align}
\end{subequations}
Hence, the baryon asymmetry is given by
\begin{equation}
n_B=n(Q,\overline{L})-n(L,\overline{Q})=\frac{g^2f^2}{2\pi^2}\int_0^{+\infty}\mathcal{F}(\omega)\omega^2d\omega,
\label{eq53}
\end{equation}
with $\mathcal{F}(\omega)$ defined as
\begin{equation}
\mathcal{F}(\omega)=\int_{-\infty}^{+\infty}\!dt \,dt^\prime e^{2i\omega(t-t')}\!\left\{e^{i[\theta(t)-\theta(t')]}-e^{i[\theta(t')-\theta(t)]}\!\right\}\!.
\label{eq54}
\end{equation}
Since inflatonic small oscillations around of the minimum of its potential are here considered, the exponential functions in $\mathcal{F}(\omega)$ can be expanded around $\theta=0$ up to second order. In this way we obtain
\begin{equation}
\mathcal{F}(\omega)=2\text{Re}\bigg{\{}\frac{\tilde{\theta}(2\omega)\tilde{\theta}^{2*}(2\omega)}{i}\bigg{\}},
\label{eq55}
\end{equation}
where $\tilde{\theta}(2\omega)$ and $\tilde{\theta}^2(2\omega)$ are given by
\begin{subequations}
\begin{align}
&\tilde{\theta}(2\omega)=\int_{-\infty}^{+\infty}dt e^{2i\omega t}\theta(t),
\label{eq56}\\
&\tilde{\theta}^2(2\omega)=\int_{-\infty}^{+\infty}dt e^{2i\omega t}\theta^2(t).
\label{eq57}
\end{align}
\end{subequations}

Let $\tilde{\theta}_0(2\omega)$ and $\tilde{\theta}^{2*}_0(2\omega)$ be the zero-order of $\tilde{\theta}(2\omega)$ and $\tilde{\theta}^{2*}(2\omega)$, respectively, as found in Ref.~\cite{Dolgov:1996qq}, and let $\mathcal{F}_0(\omega)$ be the corresponding zero-order expression for $\mathcal{F}(\omega)$.
The baryon asymmetry from $\mathcal{F}(\omega)=\mathcal{F}_0(\omega)$ furnishes the result in Eq.~(\ref{eq3}), while the first order correction, due to the fact that we use Eq.~(\ref{eq48}) in lieu of Eq.~(\ref{eq7}), provides
\begin{subequations}
\begin{align}
\nonumber
&\tilde{\theta}(2\omega)=\tilde{\theta}_0(2\omega)+\frac{\Xi\theta_I}{12i}\bigg{[}-\frac{2}{2i\omega-\frac{\Gamma}{2}+i\Omega}+\frac{2}{2i\omega-\frac{\Gamma}{2}-i\Omega}\\
&+\frac{1}{2i\omega-\Gamma+2i\Omega}-\frac{1}{2i\omega-\Gamma-2i\Omega}\bigg{]}, \label{eq58}\\
\nonumber
&\tilde{\theta}^{2*}(2\omega)=\tilde{\theta}_0^{2*}(2\omega)+\frac{\Xi\theta_I^2}{12i}\bigg{[}-\frac{2}{-2i\omega-\Gamma+2i\Omega}+\\
\nonumber
&+\frac{2}{-2i\omega-\Gamma-2i\Omega}+\frac{1}{-2i\omega-\frac{3\Gamma}{2}+3i\Omega}+\\
\nonumber
&-\frac{1}{-2i\omega-\frac{3\Gamma}{2}-3i\Omega}+\frac{1}{-2i\omega-\frac{3\Gamma}{2}+i\Omega}+\\
&-\frac{1}{-2i\omega-\frac{3\Gamma}{2}-i\Omega}\bigg{]}.
\label{eq59}
\end{align}
\end{subequations}
The full calculations can be found in Appendix~\ref{app2a}.

\subsection{The corrected baryonic asymmetry}

The baryon asymmetry can be now computed via Eqs.~(\ref{eq53}), (\ref{eq55}), and (\ref{eq58})--(\ref{eq59}). We expect a first order correction with respect to the result obtained in the background model. Because $g\ll1$, we compute the baryon asymmetry in the limit of $\Omega\gg\Gamma$. Full calculations can be found in Appendix \ref{app2b}. We obtain
\begin{equation}
n_B\!=\!\frac{g^2f^2\theta_I^3}{16\pi}\bigg{[}\Omega\!+\!\frac{49\xi\Upsilon\theta_I}{6\pi}\!\lim_{\omega\to+\infty}\!\ln\bigg{(}\frac{\omega}{\Omega}\bigg{)}\bigg{]}\!-\!\frac{\xi f^2\Upsilon\theta_I^4}{24}.
\label{eq60}
\end{equation}
The term in square brackets diverges if $\Omega$ remains finite.
However, as in Eq.~(\ref{eq6}), the inflaton mass $\Omega$ can be taken infinitely large and opportunely renormalized in order to obtain a finite quantity in square brackets. Therefore, we can perform a further renormalizion of the inflaton mass
\begin{equation}
M_{\theta}=\lim_{\Omega\to+\infty}\bigg{[}\Omega+\frac{49\xi\Upsilon\theta_I}{6\pi}\lim_{\omega\to+\infty}\ln\bigg{(}\frac{\omega}{\Omega}\bigg{)}\bigg{]}.
\label{eq61}
\end{equation}
Such a mass is not finite for any value of $\xi$. To see this, in Eq.~\eqref{eq61} we define $x=49\xi\Upsilon\theta_I/6\pi$ and then we have
\begin{equation}
\ln\bigg{(}\frac{\omega}{\Omega}\bigg{)}=\frac{M_{\theta}}{x}-\frac{\Omega}{x}\implies \omega=e^{M_{\theta}/x}\bigg{(}\Omega e^{-\Omega/x}\bigg{)}.
\label{eq62bis}
\end{equation}
We see that the limits $\Omega\to+\infty$ and $\omega\to+\infty$ are compatible if and only if $x<0$ (or $\xi<0$). Thus, to enable the renormalization, from now on we consider $\xi<0$.

Physically, the sign of $\xi$ is determinant in establishing whether the inflaton mass increases or not in presence of the non-minimal coupling.
To enhance the generated baryon asymmetry, the inflaton mass should increase, thereby increasing the likelihood of decays into fermion–antifermion pairs. Looking at Eq.~\eqref{eq14}, we notice that for $\xi<0$ the inflaton mass increases only if $R<0$. According to Eq.~\eqref{eq37}, $R$ depends upon $\dot \theta$ and, since the inflaton is an oscillating and decaying field, on average we have that $\dot{\theta}<0$. This yields a negative Ricci curvature, that is exactly what we desire to produce an increase of the inflaton mass and ensure the baryogenesis process to be highly enhanced.

Once having the renormalized $M_{\theta}$, it can substitute $\Omega$ everywhere and, as for $\Omega$, we impose $M_{\theta}\sim 10^{-6}f$. Then, with this prescription, Eq.~\eqref{eq60} becomes
\begin{equation}
n_B=\frac{g^2f^2}{16\pi}\theta_I^3M_{\theta}-\frac{\xi f^2}{24}\Upsilon\theta_I^4=n_B^{(0)}\bigg{(}1-\frac{2\pi\Xi}{3g^2}\bigg{)},
\label{eq62}
\end{equation}
where $n_B^{(0)}$ is given by Eq.~\eqref{eq3} with $\Omega$ replaced by $M_{\theta}$. Being $\Xi<0$, we obtain a positive correction that becomes relevant when $|\Xi|/g^2\sim |\xi|\theta_If/(g^2M_\theta)\sim 10^6|\xi|>1$. At perturbative level, we have $|\Xi|\sim 10^6|\xi|g^2\ll1$. Since $g^2\ll1$, having $10^6|\xi|\ge1$ is definitely plausible and cannot be excluded \emph{a priori}. Consequently, the correction to the baryon asymmetry that we obtain due to the non-minimal coupling between gravity and inflaton would turn out to be significant.

\section{Mass mixing and consequences of cosmic expansion}\label{sec:level6}

The fermionic fields $Q$ and $L$ are not mass eigenstates of the Lagrangian in Eq.~\eqref{eq8} thus, if do not decay immediately into stable lighter mass particles with appropriate quark quantum numbers, a particle which is produced as $Q$ may later rotate into a lepton $L$ and viceversa according to the weel-know phenomenon of the \emph{mass mixing}.
This possibility might be taken into account in the calculation of the baryon asymmetry.

The non-minimal coupling term does not affect the Lagrangian part that describes fermions. Hence, the mass eigenstates and the corresponding eigenvalues are the same as in the case of $\xi=0$. Consequently, the correction that arises from the mass mixing is still $[(1-\epsilon^2)/(1+\epsilon^2)]^2$, as reported at the beginning of this work.

Conventionally, the baryon asymmetry is normalized with the entropy density $s=(2\pi^2/45)g_*T_{\text{RH}}^3$ of the  reheating epoch, i.e. when the baryogenesis takes place, at a temperature $T_{\text{RH}}=0.2(200/g_*)^{1/4}\sqrt{M_{\text{Pl}}\Gamma}$ with  $g^\star$ being the effective numbers of relativistic degrees from all particles in thermal equilibrium with photons.

Therefore, including the mass mixing, we end up with
\begin{equation}
\begin{split}
&\eta=\frac{n_B}{s}=\frac{45 n_B^{(0)}}{2\pi^2g_*T_{\text{RH}}^3} \bigg{(}1-\frac{2\pi\Xi}{3g^2}\bigg{)}\bigg{(}\frac{1-\epsilon^2}{1+\epsilon^2}\bigg{)}^2.
\end{split}
\label{eq64}
\end{equation}
It is worth noticing that, in Eq.~\eqref{eq64}, $\Xi\propto\theta_I$, thus its quantitative evaluation is mandatory.
Focusing on the regime $H\le \Gamma$ and assuming that the baryon asymmetry production started when $H=\Gamma$, we obtain the definition $\theta_I=\sqrt{3/(4\pi)}\Gamma M_{\text{Pl}}/(fM_{\theta})\sim g^2$ given in Sec.~\ref{sec:level2}.

By computing explicitly Eq.~\eqref{eq64}, we obtain
\begin{equation}
\eta\simeq 10^{-4}\frac{g^5}{g_*^{1/4}}\left(\frac{M_{\text{Pl}}^3}{f^2 M_\theta}\right)^{\frac{1}{2}}\bigg{(}1-\frac{2\pi\Xi}{3g^2}\bigg{)}\biggl(\frac{1-\epsilon^2}{1+\epsilon^2}\biggl)^2.
\label{eq65}
\end{equation}

The above result neglects the particle production initiated much earlier, when $H\sim M_{\theta}$ and the inflaton field began to oscillate around the bottom of the potential.
Thus, to include earlier particle production, we indicate with  $\theta_I^\prime$ and $\Xi^\prime$ the values of $\theta$ and $\Xi$ obtained for $H=M_{\theta}$, respectively.
We find that $\theta_I'/\theta_I=M_{\theta}/\Gamma$ and $\Xi'=(M_{\theta}/\Gamma)\Xi\sim \Xi/g^2\gg\Xi$.
This implies that in Eq.\eqref{eq65} the zero order $n_B^{(0)}$, which is proportional to $\theta_I^3$, has to be corrected with a multiplicative factor $(\theta_I'/\theta_I)^3=M^3_{\theta}/\Gamma^3$, whereas a factor $M_{\theta}/\Gamma$ appears in front of $\Xi$.

However, when we consider the particle production to start at $H=M_{\theta}$, we need to include the expansion of the universe. Between the time of peak production of baryon asymmetry at $t_a\sim1/M_{\theta}$ and the peak entropy production at $t_b\sim1/\Gamma$, the baryon asymmetry is diluted by a factor of $(a_a/a_b)^3\sim(t_a/t_b)^2\sim(\Gamma/M_{\theta})^2$ due to the expansion of the universe \cite{Dolgov:1996qq}. In this estimate, we are considering the universe to be matter dominated, with $a\propto t^{2/3}$ during reheating. Therefore, Eq.~\eqref{eq65} must be modified by a total factor $(M_{\theta}/\Gamma)^3(\Gamma/M_{\theta})^2=M_{\theta}/\Gamma=8\pi/g^2$. The same factor appears also in front of $\Xi$.

Modifying accordingly Eq.~\eqref{eq65}, we finally obtain
\begin{equation}
\eta\simeq 3\times10^{-3}\frac{g^3}{g_*^{1/4}}\!\bigg{(}\!\frac{M_{\text{Pl}}^2}{f^2 M_\theta}\!\bigg{)}^{\!\frac{1}{2}}\!\left(\!1-\frac{16\pi^2\Xi}{3g^4}\!\right)\!\biggl(\!\frac{1-\epsilon^2}{1+\epsilon^2}\!\biggl)^2\!,
\label{eq66}
\end{equation}
that is modified by a corrective factor $1-16\pi^2\Xi/(3g^4)$ with respect to the zero order result.

\begin{figure*}
\centering
{\hfill
\includegraphics[width=0.565\hsize,clip]{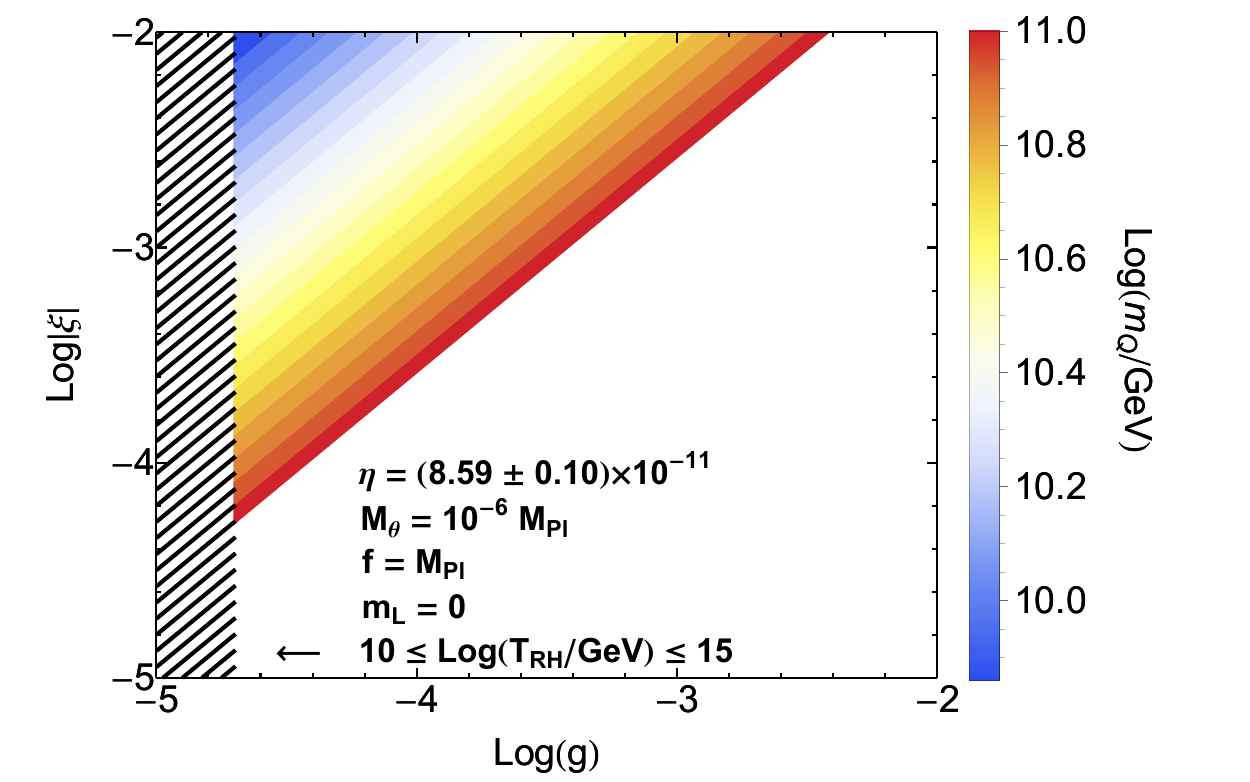}
\hfill
\includegraphics[width=0.425\hsize,clip]{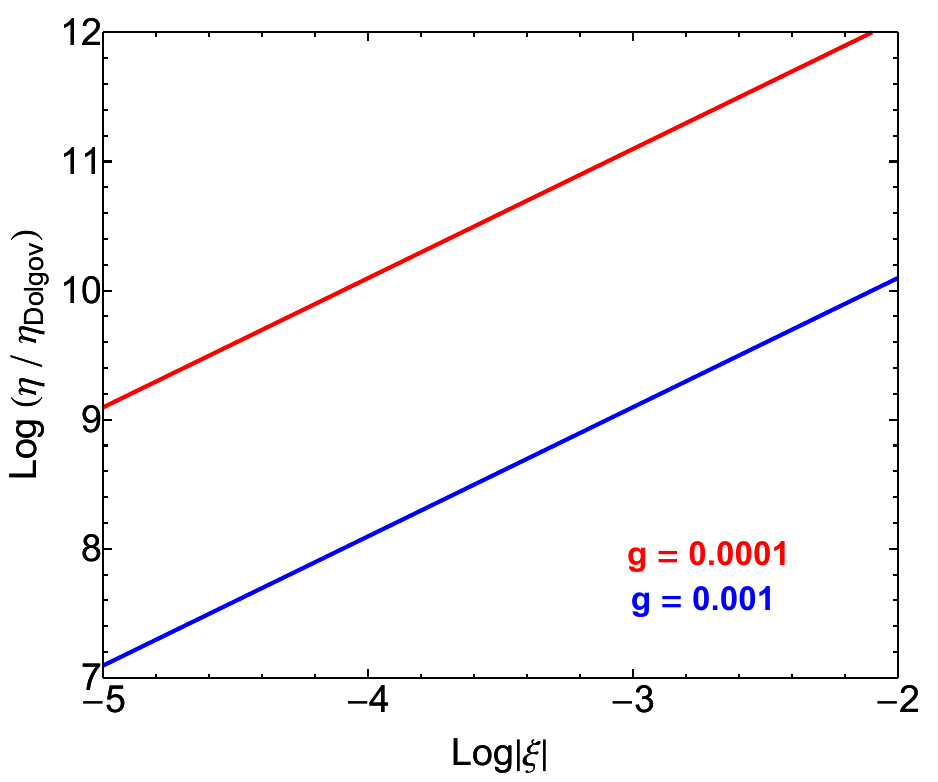}
\hfill}
\caption{\emph{Left}: contour plot of the allowed region of the $\log{(g)}-\log{|\xi|}$ plane, compatible with the observational constraint $\eta=(8.59\pm0.10)\times 10^{-11}$. The chosen parameters are summarized in the legend. \emph{Right}: ratio between our $\eta$ from Eq.~\eqref{eq66} and the background result \cite{Dolgov:1996qq} as function of $\xi$, for $g=0.0001$ and $g=0.001$.}
\label{fig2}
\end{figure*}

The inclusion of the expansion of the universe greatly increases our correction with respect to the background model. More precisely, this correction is of the order of $|\Xi|/g^4\sim 10^6|\xi|/g^2$, enhanced by a factor $1/g^2$ with respect to the case in which the expansion is neglected.

The enhancement of the baryogenesis due to the non-minimal coupling is physically justified by the fact that such a term increases the mass of the inflaton by $\xi R$, that is a positive quantity. The Ricci curvature in a FLRW spacetime has the general form $R=-6(\ddot{a}/a+H^2)$ that, specialized for a matter-dominated reheating -- when baryogenesis is assumed to occur -- becomes $R=-3H^2$. Thus, considering that the expansion of the universe increases the Ricci curvature and consequently the first-order correction due to the non-minimal coupling, the baryon asymmetry becomes relevant if $10^6|\xi|$ is comparable with $g^2$.
This fact turns out to be very likely. Indeed, in discussing the result obtained without considering the expansion of the universe, we say that having $10^6|\xi|\ge1$ cannot be excluded \emph{a priori}. Even more so, including the expansion effects, the condition $10^6|\xi|/g^2\ge 1$ is satisfied.

The contour plot in Fig.~\ref{fig2} shows the allowed region of the $\log{(g)}-\log{|\xi|}$ plane, provided that the normalized baryon asymmetry written in Eq.~\eqref{eq66} results to be compatible with its experimental value $\eta=(8.59\pm0.10)\cdot 10^{-11}$ and $T_{\rm RH}$ bounded to be in the range $10^{10}$--$10^{15}$~GeV. For simplicity, we consider $m_L=0$ and a variable $m_Q$, always taken much smaller with respect to the inflaton mass $M_{\theta}$.
As an important remark, the allowed values of $\xi$ and $g$ that we find depend on our free choice of taking $f/\Lambda\sim 10^3$. For other choices of this ratio, the values of the coupling constants compatible with the experimental baryon asymmetry would change.

In Fig.~\ref{fig2}, we also show the normalized baryon asymmetry from Eq.~\eqref{eq66} divided by the standard result \cite{Dolgov:1996qq}, as function of $\xi$, for $g=0.0001$ and $g=0.001$.
Accordingly, we find corrections that are larger than the corresponding background results as much as $g$ is small enough, permitting larger enhancements of baryogenesis, in agreement with Planck observations \cite{Planck:2018jri}.

\section{Final remarks}\label{sec:level7}

In this work, we studied the mechanism of spontaneous baryogenesis in producing the baryon asymmetry.
We extended the standard approach including a Yukawa-like contribution that couples the Ricci scalar with the inflaton field. Accordingly, we emphasized that the non-minimal coupling could increase the inflaton mass and consequently the probability for its decay into fermion-antifermion pairs. The overall process therefore enhances the baryogenesis itself and shows that, at the energy scales here involved, there is no apparent reason to neglect \emph{a priori} non-minimal couplings with curvature.

In so doing, we did not include additional non-minimal couplings between fermionic and gravitational sectors, which were left unmodified. Instead, we introduced the non-minimal coupling only among classical fields, motivated by the fact that inflationary models with non-minimal couplings are currently supported by Planck results \cite{Planck:2018jri}, favoring potentials where the coupling between inflaton and spacetime curvature plays a central role.
An additional advantage of this setup was the absence of mass mixing terms arising from quadratic couplings with fermionic fields. By quantizing only the fermionic sector, while treating inflaton and gravitational fields classically, we avoided such complications and remained consistent with the original framework of spontaneous baryogenesis.

We derived the equation of motion for the inflaton and obtained a first-order perturbative solution in the non-minimal coupling constant $\xi$. The baryon asymmetry was then computed following the strategy of Ref.~\cite{Dolgov:1996qq}, with a crucial difference: instead of using only the zeroth-order solution for the inflaton, we included the resulting first-order correction as well. This led to a modified expression for the baryon asymmetry, differing from that in Ref. \cite{Dolgov:1996qq} by a multiplicative factor $1 - 16\pi^2 \Xi/(3g^4)$, which proved to be particularly significant in the regime of non-negligible expansion of the universe. Additionally, we showed that the inflaton mass received a correction due to the non-minimal coupling, effectively resulting in its renormalization. This requirement forced the coupling constant $\xi$ to be negative, since a positive value would have led to an ill-defined mass shift, implying a positive correction in the baryon asymmetry.

Future works will aim to introduce further extensions into the model, e.g. like considering the coupling between electromagnetic field and inflaton that, in principle, may affect the production of baryon asymmetry. Another possibility could be the introduction of a disformal coupling between inflaton and gravity proportional to $\partial_{\mu}\theta \partial_{\nu}\theta R^{\mu\nu}$, given that in this work we have studied the conformal case. Finally, in order to make the model more physical, quantum chromodynamics must be definitely introduced, since the field $Q$ considered in the model is not a real quark, because the color number is not included.

\section*{Acknowledgments}
MD is thankful to Youri Carloni and Tommaso Mengoni for insightful comments on early-time universe. OL acknowledges support by the  Fondazione  ICSC, Spoke 3 Astrophysics and Cosmos Observations. National Recovery and Resilience Plan (Piano Nazionale di Ripresa e Resilienza, PNRR) Project ID $CN00000013$ ``Italian Research Center on  High-Performance Computing, Big Data and Quantum Computing" funded by MUR Missione 4 Componente 2 Investimento 1.4: Potenziamento strutture di ricerca e creazione di ``campioni nazionali di R\&S (M4C2-19)" - Next Generation EU (NGEU). OL is also grateful to Aniello Quaranta for interesting discussions on the topic of baryogenesis. MM acknowledges support from the project OASIS, ``PNRR Bando a Cascata da INAF M4C2 - INV. 1.4''.

\appendix

\section{Dirac equation in FLRW spacetime}\label{app0}

The free Dirac equation in FLRW spacetime is
\begin{widetext}
\begin{equation}
\begin{split}
&i\gamma^{\mu}\nabla_{\mu}\psi-m\psi=0\implies ie^{\mu}_a\gamma^{a}(\partial_{\mu}+\Omega_{\mu})\psi-m\psi=0\implies \\&\implies i\gamma^0\dot{\psi}+\frac{i}{a}\gamma^i\partial_i\psi+i\gamma^i\frac{\dot{a}\gamma_i\gamma_0}{2a}=m\psi\implies i\gamma^0\bigg{(}\dot{\psi}+\frac{3H}{2}\psi\bigg{)}+i\gamma^i\frac{\partial_i\psi}{a}=m\psi,
\end{split}
\label{eqAPP01}
\end{equation}
\end{widetext}
It is convenient to define $\tilde{\psi}$ such that $\psi=a^{3/2}\tilde{\psi}$. For the new field $\tilde{\psi}$, we obtain
\begin{widetext}
\begin{equation}
i\gamma^0\bigg{(}-\frac{3}{2}a^{-3/2}H\tilde{\psi}+a^{-3/2}\dot{\tilde{\psi}}+\frac{3}{2}Ha^{-3/2}\tilde{\psi}\bigg{)}+a^{-3/2}i\gamma^i\frac{\partial_i\tilde{\psi}}{a}=ma^{-3/2}\tilde{\psi}\implies i\gamma^0\dot{\tilde{\psi}}+i\gamma^i\frac{\partial_i\tilde{\psi}}{a}=m\tilde{\psi}.
\label{eqAPP02}
\end{equation}
\end{widetext}
Eq.~\eqref{eqAPP02} is the free Dirac equation with the spatial components of the 4-momentum rescaled by a factor $1/a$. In the regime of negligible expansion of the universe\footnote{In this phase, we are neglecting the expansion of the universe.}, $a$ can be considered just a parameter to rescale. Hence, the canonically quantized solution of Eq.~\eqref{eqAPP02} is the same as in Minkowski spacetime but with the momentum $\vec{p}$ rescaled by a factor $1/a$, that is
\begin{widetext}
\begin{equation}
\tilde{\psi}(t,\vec{x})=\sum_s\int\frac{d^3p'}{(2\pi)^3}\frac{1}{\sqrt{2E'_p}}\bigg{(}u(\vec{p'},s)\hat a(\vec{p'},s)e^{-i(E'_pt-\vec{p'}\cdot \vec{x})}+v(\vec{p'},s)\hat b^{\dag}(\vec{p'},s)e^{i(E'_pt-\vec{p'}\cdot\vec{x})}\bigg{)},
\label{eqAPP03}
\end{equation}
\end{widetext}
where $\vec{p'}=\vec{p}/a$ and $E'^2_p=p'^2+m^2=p^2/a^2+m^2$. By a rescaling of the 3D-momentum $\vec{p'}\to\vec{p}=a\vec{p'}$, we get back for $\tilde{\psi}$ the canonical quantization of the Dirac field in Minkowski spacetime. This should not be surprising, because summing over all the three dimensional momenta makes the Dirac field be independent of any rescaling of $\vec{p}$. For $\psi$, thus, we obtain
\begin{widetext}
\begin{equation}
\psi(t,\vec{x})=a^{-3/2}(t)\sum_s\int\frac{d^3p}{(2\pi)^3}\frac{1}{\sqrt{2E_p}}\bigg{(}u(\vec{p},s)\hat a(\vec{p},s)e^{-i(E_pt-\vec{p}\cdot \vec{x})}+v(\vec{p},s)\hat b^{\dag}(\vec{p},s)e^{i(E_pt-\vec{p}\cdot\vec{x})}\bigg{)}.
\label{eqAPP03}
\end{equation}
\end{widetext}

\section{VEV computations}\label{app1}

In this appendix, we show the calculations leading to Eq.~\eqref{eq26}--\eqref{eq27}.

\subsection{Calculation of $\braket{\overline{Q}L+\overline{L}Q}$}\label{app1a}

To compute $\braket{\overline{Q}L+\overline{L}Q}$ we make use of the calculation for $\braket{\nabla_{\mu}J^{\mu}}$ shown in Ref.~\cite{Dolgov:1994zq}. Indeed, we have
\begin{widetext}
\begin{equation}
\begin{split}
&\braket{\nabla_{\mu}J^{\mu}}=\frac{igf}{\sqrt{2}}\braket{\overline{Q}L-\overline{L}Q}=a^{-3}(t)\frac{g^2f^2}{2\pi^2} \bigg{[}\int_{-\infty}^{t_x} dt_y\bigg{(}e^{i\theta(t_y)-i\theta(t_x)}-e^{-i\theta(t_y)+i\theta(t_x)}\bigg{)}\int_0^{+\infty} d\omega \omega^2e^{-2i\omega(t_x-t_y)}+h.c.\bigg{]}
\implies \\&\implies \frac{igf}{\sqrt{2}}\braket{\overline{Q}L}=a^{-3}(t)\frac{g^2f^2}{2\pi^2}\int_{-\infty}^{t_x} dt_y\bigg{(}e^{i\theta(t_y)-i\theta(t_x)}-e^{-i\theta(t_y)+i\theta(t_x)}\bigg{)} \int_0^{+\infty} d\omega \omega^2e^{-2i\omega(t_x-t_y)}\implies \\&\implies \frac{igf}{\sqrt{2}}\braket{\overline{Q}L+\overline{L}Q}=\bigg{\{}a^{-3}(t)\frac{g^2f^2}{2\pi^2}\int_{-\infty}^{t_x} dt_y\bigg{(}e^{i\theta(t_y)-i\theta(t_x)}-e^{-i\theta(t_y)+i\theta(t_x)}\bigg{)} \int_0^{+\infty} d\omega \omega^2e^{-2i\omega(t_x-t_y)}\bigg{\}}-h.c.=\\&=a^{-3}(t)\frac{g^2f^2}{2\pi^2}\int_{-\infty}^{t_x}dt_y\bigg{(} e^{i\theta(t_y)-i\theta(t_x)}-e^{-i\theta(t_y)+i\theta(t_x)}\bigg{)} \int_0^{+\infty}d\omega \omega^22\cos(2\omega(t_x-t_y))=\\&=a^{-3}(t)\frac{2ig^2f^2}{\pi^2}\int_{-\infty}^0\sin[\theta(t'+t_x)-\theta(t_x)]\int_0^{+\infty}d\omega \omega^2\cos(2\omega t').
\end{split}
\label{eqAPP1}
\end{equation}
\end{widetext}
Proceeding the same way as in Ref.~\cite{Dolgov:1994zq}, Eq.~\eqref{eqAPP1} yields
\begin{widetext}
\begin{equation}
\begin{split}
&a^3(t)\frac{gf}{\sqrt{2}}\braket{\overline{Q}L+\overline{L}Q}=\frac{2g^2f^2}{\pi^2}\int_{-\infty}^0dt'\sin[\theta(t+t')-\theta(t)]\int_0^{+\infty}d\omega\omega^2\cos(2\omega t')=\\&=-\frac{g^2f^2}{4\pi^2}\lim_{\omega\to+\infty}\int_{-\infty}^0\frac{\partial^2\sin(\Delta\theta)}{\partial t'^2}\frac{\sin(2\omega t')}{t'}dt'=-\frac{g^2f^2}{4\pi^2}\lim_{\omega\to+\infty}\int_{-\infty}^0\frac{\sin(2\omega t')}{t'}[\ddot{\theta}(t+t')\cos(\Delta\theta)-\dot{\theta}^2(t+t')\sin(\Delta \theta)]dt'\simeq \\&\simeq \frac{g^2f^2\Omega^2}{4\pi^2}\theta_i(t)\lim_{\omega\to+\infty}\int_{-\infty}^0\frac{\sin(2\omega t')}{t'}\cos[\Omega(t+t')]dt'=\\&=
\frac{g^2f^2\Omega^2}{4\pi^2}\theta_i(t)\frac{\pi}{2}\cos(\Omega t)+\frac{g^2f^2\Omega^2}{4\pi^2}\theta_i(t)\frac{1}{2}\sin(\Omega t)\lim_{\omega\to+\infty}\bigg{[}\int_0^{+\infty}\frac{\cos(2\omega t'-\Omega t')}{t'}dt'-\int_0^{+\infty}\frac{\cos(2\omega t'+\Omega t')}{t'}dt'\bigg{]}=\\&=\frac{g^2f^2\Omega^2}{8\pi}\theta(t)+\frac{g^2f^2\Omega^2}{8\pi^2}\dot{\theta}(t)\lim_{\omega\to+\infty}\bigg{[}\text{Ci}\bigg{(}2\omega t'-\Omega t'\bigg{)}-\text{Ci}\bigg{(}2\omega t'+\Omega t'\bigg{)}\bigg{]}\bigg{|}_0^{+\infty}=
\frac{g^2f^2\Omega^2}{8\pi}\theta(t)=\Gamma\Omega f^2\theta(t).
\end{split}
\label{eqAPP2}
\end{equation}
\end{widetext}

\subsection{\label{app1b}Calculation of $\braket{J^0}$}

\begin{widetext}
\begin{equation}
\begin{split}
&\braket{J^0}=\braket{\overline{Q}_0\gamma^0Q_0}=a^{-3}(t)\sum_{s,r}\int \frac{d^3p}{(2\pi)^3}\frac{1}{\sqrt{2E_p}}\int \frac{d^3q}{(2\pi)^3}\frac{1}{\sqrt{2E_q}}\bigg{\langle}\bigg{(}\overline{u}(\vec{p},s)\hat a^{\dag}(\vec{p},s)e^{i(E_pt-\vec{p}\cdot\vec{x})}+\overline{v}(\vec{p},s)\hat b(\overline{p},s)e^{-i(E_pt-\vec{p}\cdot\vec{x})}\bigg{)}\times \\
&\gamma^0\bigg{(}u(\vec{q},r)\hat a(\vec{q},r)e^{-i(E_qt-\vec{q}\cdot\vec{x})}+v(\vec{q},r)\hat b^{\dag}(\overline{q},r)e^{i(E_qt-\vec{q}\cdot\vec{x}))}\bigg{)}\bigg{\rangle}=\\&=a^{-3}(t)\sum_{s,r}\int \frac{d^3p}{(2\pi)^3}\frac{1}{\sqrt{2E_p}}\int \frac{d^3q}{(2\pi)^3}\frac{1}{\sqrt{2E_q}}\overline{v}(\vec{p},s)\gamma^0v(\vec{q},r)\times  e^{-i(E_pt-\vec{p}\cdot\vec{x})}e^{i(E_qt-\vec{q}\cdot\vec{x})}(2\pi)^3\delta^{(3)}(\overline{p}-\overline{q})\delta_{s,r}=\\&=a^{-3}(t)\sum_s\int \frac{d^3p}{(2\pi)^3}\frac{1}{2E_p}\overline{v}_a(\vec{p},s)\gamma^0_{ab}v_b(\vec{p},s)=a^{-3}(t)\int \frac{d^3p}{(2\pi)^3}\frac{1}{2E_p}\cancel{p}_{ba}\gamma^0_{ab}=
\frac{P^3_{\text{max}}}{3\pi^2}a^{-3}(t).
\end{split}
\label{eqAPP3}
\end{equation}
\end{widetext}

\section{Generated baryon asymmetry}\label{app2}

We here show the calculations for $\tilde{\theta}(2\omega)$ and $\tilde{\theta}^{2*}(2\omega)$ of Eq.~\eqref{eq58}--\eqref{eq59}. Moreover, we show some steps that lead to the result for the baryon asymmetry of Eq.~\eqref{eq60}.

\subsection{Calculation of $\tilde{\theta}(2\omega)$ and $\tilde{\theta}^{2*}(2\omega)$}\label{app2a}

For $\tilde{\theta}(2\omega)$ and $\tilde{\theta}^{2*}(2\omega)$, we have, respectively
\begin{widetext}
\begin{align}
\nonumber
\tilde{\theta}(2\omega)=&\,\theta_I\int_{-\infty}^0dte^{2i\omega t}+\theta_I\int_0^{+\infty}dte^{2i\omega t}e^{-\frac{\Gamma t}{2}}\bigg{[}\cos(\Omega t)+\frac{\Xi}{3}\sin(\Omega t)\bigg{(}1-e^{-\frac{\Gamma t}{2}}\cos(\Omega t)\bigg{)}\bigg{]}=\\
&=
\tilde{\theta}_0(2\omega)+\frac{\Xi\theta_I}{12i}\bigg{[}-\frac{2}{2i\omega-\frac{\Gamma}{2}+i\Omega}+\frac{2}{2i\omega-\frac{\Gamma}{2}-i\Omega}+\frac{1}{2i\omega-\Gamma+2i\Omega}-\frac{1}{2i\omega-\Gamma-2i\Omega}\bigg{]}. \label{eqAPP4}\\
\nonumber
\tilde{\theta}^{2*}(2\omega)=&\,\theta_I^2\int_{-\infty}^0e^{-2i\omega t}dt+\theta_I^2\int_0^{+\infty}e^{-2i\omega t}e^{-\Gamma t}\bigg{[}\cos^2(\Omega t)+\frac{2}{3}\Xi\cos(\Omega t)\sin(\Omega t)\bigg{(}1-e^{-\frac{\Gamma t}{2}}\cos(\Omega t)\bigg{)}\bigg{]}dt=\\
\nonumber
&=
\tilde{\theta}_0^{2*}(2\omega)+\frac{\Xi\theta_I^2}{12i}\bigg{[}-\frac{2}{-2i\omega-\Gamma+2i\Omega}+\frac{2}{-2i\omega-\Gamma-2i\Omega}+\frac{1}{-2i\omega-\frac{3\Gamma}{2}+3i\Omega}-\frac{1}{-2i\omega-\frac{3\Gamma}{2}-3i\Omega}+\\
&+\frac{1}{-2i\omega-\frac{3\Gamma}{2}+i\Omega}-\frac{1}{-2i\omega-\frac{3\Gamma}{2}-i\Omega}\bigg{]}.
\label{eqAPP7}
\end{align}
\end{widetext}

\subsection{\label{app2b}Calculation of the baryon asymmetry}

With the expressions for $\tilde{\theta}(2\omega)$ and $\tilde{\theta}^{2*}(2\omega)$ obtained above, we can compute $\mathcal{F}(\omega)$ via Eq. (\ref{eq55}). We obtain
\begin{widetext}
\begin{equation}
\begin{split}
&\mathcal{F}(\omega)=2\text{Re}\bigg{\{}\frac{\tilde{\theta}(2\omega)\tilde{\theta}^{2*}(2\omega)}{i}\bigg{\}}
=\mathcal{F}_0(\omega)-\frac{\Xi\theta_I}{6}\text{Re}\bigg{\{}\theta_I\tilde{\theta}_0(2\omega)\bigg{[}-\frac{2}{-2i\omega-\Gamma+2i\Omega}+\frac{2}{-2i\omega-\Gamma-2i\Omega}+\frac{1}{-2i\omega-\frac{3\Gamma}{2}+3i\Omega}+\\&-\frac{1}{-2i\omega-\frac{3\Gamma}{2}-3i\Omega}+\frac{1}{-2i\omega-\frac{3\Gamma}{2}+i\Omega}-\frac{1}{-2i\omega-\frac{3\Gamma}{2}-i\Omega}\bigg{]}+\\&+\tilde{\theta}_0^{2*}(2\omega)\bigg{[}-\frac{2}{2i\omega-\frac{\Gamma}{2}+i\Omega}+\frac{2}{2i\omega-\frac{\Gamma}{2}-i\Omega}+\frac{1}{2i\omega-\Gamma+2i\Omega}-\frac{1}{2i\omega-\Gamma-2i\Omega}\bigg{]}\bigg{\}}.
\end{split}
\label{eqAPP11}
\end{equation}
\end{widetext}
The baryon asymmetry is computed by Eq.~\eqref{eq53}, with $\mathcal{F}(\omega)$ given by the expression above. Let $n_B^{(0)}$ be the baryon asymmetry given by Eq.~\eqref{eq3}, we have
\begin{widetext}
\begin{equation}
n_B=\frac{g^2f^2}{2\pi^2}\int_0^{+\infty}\omega^2\mathcal{F}(\omega)d\omega=n_B^{(0)}-
\frac{g^2f^2}{2\pi^2}\frac{\Xi\theta_I}{6}\int_0^{\infty}\omega^2\mathcal{G}(\omega),
\label{eqAPP12}
\end{equation}
\end{widetext}
where $\mathcal{G}(\omega)$ is real part of the quantity in curly brackets in Eq.~\eqref{eqAPP11}. The correction to $n_B^{(0)}$ can be evaluated by computing the integral in Eq.~\eqref{eqAPP12} in the limit of $\Omega\gg\Gamma$. It results
\begin{widetext}
\begin{equation}
\begin{split}
&n_B=n_B^{(0)}-\frac{g^2f^2\Xi\theta_I^3}{48\pi^2}\bigg{[}\frac{\Omega^2\pi}{4\Gamma}-\frac{49}{2}\Omega\lim_{\omega\to+\infty}\ln\bigg{(}\frac{\omega}{\Omega}\bigg{)}\bigg{]}=\\&=\frac{1}{16\pi}\Omega g^2f^2\theta_I^3+\frac{49}{96\pi^2}\Omega g^2f^2\Xi\theta_I^3\lim_{\omega\to+\infty}\ln\bigg{(}\frac{\omega}{\Omega}\bigg{)}-\frac{1}{192\pi}\frac{\Omega}{\Gamma}\Omega g^2f^2\Xi\theta_I^3=\\&=\frac{1}{16\pi}\Omega g^2f^2\theta_I^3\bigg{[}1+\frac{49\Xi}{6\pi}\lim_{\omega\to+\infty}\ln\bigg{(}\frac{\omega}{\Omega}\bigg{)}\bigg{]}-\frac{1}{24}\Omega f^2\theta_I^3\Xi=\frac{1}{16\pi}g^2f^2\theta_I^3\bigg{[}\Omega+\frac{49\xi\Upsilon\theta_I}{6\pi}\lim_{\omega\to+\infty}\ln\bigg{(}\frac{\omega}{\Omega}\bigg{)}\bigg{]}-\frac{1}{24}f^2\xi\Upsilon\theta_I^4.
\end{split}
\label{eqAPP14}
\end{equation}
\end{widetext}


\begin{thebibliography}{45}%
\makeatletter
\providecommand \@ifxundefined [1]{%
 \@ifx{#1\undefined}
}%
\providecommand \@ifnum [1]{%
 \ifnum #1\expandafter \@firstoftwo
 \else \expandafter \@secondoftwo
 \fi
}%
\providecommand \@ifx [1]{%
 \ifx #1\expandafter \@firstoftwo
 \else \expandafter \@secondoftwo
 \fi
}%
\providecommand \natexlab [1]{#1}%
\providecommand \enquote  [1]{``#1''}%
\providecommand \bibnamefont  [1]{#1}%
\providecommand \bibfnamefont [1]{#1}%
\providecommand \citenamefont [1]{#1}%
\providecommand \href@noop [0]{\@secondoftwo}%
\providecommand \href [0]{\begingroup \@sanitize@url \@href}%
\providecommand \@href[1]{\@@startlink{#1}\@@href}%
\providecommand \@@href[1]{\endgroup#1\@@endlink}%
\providecommand \@sanitize@url [0]{\catcode `\\12\catcode `\$12\catcode `\&12\catcode `\#12\catcode `\^12\catcode `\_12\catcode `\%12\relax}%
\providecommand \@@startlink[1]{}%
\providecommand \@@endlink[0]{}%
\providecommand \url  [0]{\begingroup\@sanitize@url \@url }%
\providecommand \@url [1]{\endgroup\@href {#1}{\urlprefix }}%
\providecommand \urlprefix  [0]{URL }%
\providecommand \Eprint [0]{\href }%
\providecommand \doibase [0]{http://dx.doi.org/}%
\providecommand \selectlanguage [0]{\@gobble}%
\providecommand \bibinfo  [0]{\@secondoftwo}%
\providecommand \bibfield  [0]{\@secondoftwo}%
\providecommand \translation [1]{[#1]}%
\providecommand \BibitemOpen [0]{}%
\providecommand \bibitemStop [0]{}%
\providecommand \bibitemNoStop [0]{.\EOS\space}%
\providecommand \EOS [0]{\spacefactor3000\relax}%
\providecommand \BibitemShut  [1]{\csname bibitem#1\endcsname}%
\let\auto@bib@innerbib\@empty
\bibitem [{\citenamefont {Cline}(2006)}]{Cline:2006ts}%
  \BibitemOpen
  \bibfield  {author} {\bibinfo {author} {\bibfnamefont {J.~M.}\ \bibnamefont {Cline}},\ }in\ \href@noop {} {\emph {\bibinfo {booktitle} {{Les Houches Summer School - Session 86: Particle Physics and Cosmology: The Fabric of Spacetime}}}}\ (\bibinfo {year} {2006})\ \Eprint {http://arxiv.org/abs/hep-ph/0609145} {arXiv:hep-ph/0609145} \BibitemShut {NoStop}%
\bibitem [{\citenamefont {Dimopoulos}\ and\ \citenamefont {Susskind}(1979)}]{Dimopoulos:1978pw}%
  \BibitemOpen
  \bibfield  {author} {\bibinfo {author} {\bibfnamefont {S.}~\bibnamefont {Dimopoulos}}\ and\ \bibinfo {author} {\bibfnamefont {L.}~\bibnamefont {Susskind}},\ }\href {\doibase 10.1016/0370-2693(79)90366-6} {\bibfield  {journal} {\bibinfo  {journal} {Phys. Lett. B}\ }\textbf {\bibinfo {volume} {81}},\ \bibinfo {pages} {416} (\bibinfo {year} {1979})}\BibitemShut {NoStop}%
\bibitem [{\citenamefont {Cohen}\ \emph {et~al.}(1991)\citenamefont {Cohen}, \citenamefont {Kaplan},\ and\ \citenamefont {Nelson}}]{Cohen:1990it}%
  \BibitemOpen
  \bibfield  {author} {\bibinfo {author} {\bibfnamefont {A.~G.}\ \bibnamefont {Cohen}}, \bibinfo {author} {\bibfnamefont {D.~B.}\ \bibnamefont {Kaplan}}, \ and\ \bibinfo {author} {\bibfnamefont {A.~E.}\ \bibnamefont {Nelson}},\ }\href {\doibase 10.1016/0550-3213(91)90395-E} {\bibfield  {journal} {\bibinfo  {journal} {Nucl. Phys. B}\ }\textbf {\bibinfo {volume} {349}},\ \bibinfo {pages} {727} (\bibinfo {year} {1991})}\BibitemShut {NoStop}%
\bibitem [{\citenamefont {Farrar}\ and\ \citenamefont {Shaposhnikov}(1994)}]{Farrar:1993hn}%
  \BibitemOpen
  \bibfield  {author} {\bibinfo {author} {\bibfnamefont {G.~R.}\ \bibnamefont {Farrar}}\ and\ \bibinfo {author} {\bibfnamefont {M.~E.}\ \bibnamefont {Shaposhnikov}},\ }\href {\doibase 10.1103/PhysRevD.50.774} {\bibfield  {journal} {\bibinfo  {journal} {Phys. Rev. D}\ }\textbf {\bibinfo {volume} {50}},\ \bibinfo {pages} {774} (\bibinfo {year} {1994})},\ \Eprint {http://arxiv.org/abs/hep-ph/9305275} {arXiv:hep-ph/9305275} \BibitemShut {NoStop}%
\bibitem [{\citenamefont {Riotto}(1998)}]{Riotto:1998bt}%
  \BibitemOpen
  \bibfield  {author} {\bibinfo {author} {\bibfnamefont {A.}~\bibnamefont {Riotto}},\ }in\ \href@noop {} {\emph {\bibinfo {booktitle} {{ICTP Summer School in High-Energy Physics and Cosmology}}}}\ (\bibinfo {year} {1998})\ pp.\ \bibinfo {pages} {326--436},\ \Eprint {http://arxiv.org/abs/hep-ph/9807454} {arXiv:hep-ph/9807454} \BibitemShut {NoStop}%
\bibitem [{\citenamefont {Bodeker}\ and\ \citenamefont {Buchmuller}(2021)}]{Bodeker:2020ghk}%
  \BibitemOpen
  \bibfield  {author} {\bibinfo {author} {\bibfnamefont {D.}~\bibnamefont {Bodeker}}\ and\ \bibinfo {author} {\bibfnamefont {W.}~\bibnamefont {Buchmuller}},\ }\href {\doibase 10.1103/RevModPhys.93.035004} {\bibfield  {journal} {\bibinfo  {journal} {Rev. Mod. Phys.}\ }\textbf {\bibinfo {volume} {93}},\ \bibinfo {pages} {035004} (\bibinfo {year} {2021})},\ \Eprint {http://arxiv.org/abs/2009.07294} {arXiv:2009.07294 [hep-ph]} \BibitemShut {NoStop}%
\bibitem [{\citenamefont {Fong}\ \emph {et~al.}(2012)\citenamefont {Fong}, \citenamefont {Nardi},\ and\ \citenamefont {Riotto}}]{Fong:2012buy}%
  \BibitemOpen
  \bibfield  {author} {\bibinfo {author} {\bibfnamefont {C.~S.}\ \bibnamefont {Fong}}, \bibinfo {author} {\bibfnamefont {E.}~\bibnamefont {Nardi}}, \ and\ \bibinfo {author} {\bibfnamefont {A.}~\bibnamefont {Riotto}},\ }\href {\doibase 10.1155/2012/158303} {\bibfield  {journal} {\bibinfo  {journal} {Adv. High Energy Phys.}\ }\textbf {\bibinfo {volume} {2012}},\ \bibinfo {pages} {158303} (\bibinfo {year} {2012})},\ \Eprint {http://arxiv.org/abs/1301.3062} {arXiv:1301.3062 [hep-ph]} \BibitemShut {NoStop}%
\bibitem [{\citenamefont {Davidson}\ \emph {et~al.}(2008)\citenamefont {Davidson}, \citenamefont {Nardi},\ and\ \citenamefont {Nir}}]{Davidson:2008bu}%
  \BibitemOpen
  \bibfield  {author} {\bibinfo {author} {\bibfnamefont {S.}~\bibnamefont {Davidson}}, \bibinfo {author} {\bibfnamefont {E.}~\bibnamefont {Nardi}}, \ and\ \bibinfo {author} {\bibfnamefont {Y.}~\bibnamefont {Nir}},\ }\href {\doibase 10.1016/j.physrep.2008.06.002} {\bibfield  {journal} {\bibinfo  {journal} {Phys. Rept.}\ }\textbf {\bibinfo {volume} {466}},\ \bibinfo {pages} {105} (\bibinfo {year} {2008})},\ \Eprint {http://arxiv.org/abs/0802.2962} {arXiv:0802.2962 [hep-ph]} \BibitemShut {NoStop}%
\bibitem [{\citenamefont {Dey}\ \emph {et~al.}(2023)\citenamefont {Dey}, \citenamefont {Jha}, \citenamefont {Mukherjee},\ and\ \citenamefont {Sahoo}}]{Dey:2021ecr}%
  \BibitemOpen
  \bibfield  {author} {\bibinfo {author} {\bibfnamefont {U.~K.}\ \bibnamefont {Dey}}, \bibinfo {author} {\bibfnamefont {T.}~\bibnamefont {Jha}}, \bibinfo {author} {\bibfnamefont {A.}~\bibnamefont {Mukherjee}}, \ and\ \bibinfo {author} {\bibfnamefont {N.}~\bibnamefont {Sahoo}},\ }\href {\doibase 10.1088/1361-6471/ac9fe5} {\bibfield  {journal} {\bibinfo  {journal} {J. Phys. G}\ }\textbf {\bibinfo {volume} {50}},\ \bibinfo {pages} {015004} (\bibinfo {year} {2023})},\ \Eprint {http://arxiv.org/abs/2102.04494} {arXiv:2102.04494 [hep-ph]} \BibitemShut {NoStop}%
\bibitem [{\citenamefont {Qi}\ and\ \citenamefont {Sun}(2024)}]{Qi:2024pqe}%
  \BibitemOpen
  \bibfield  {author} {\bibinfo {author} {\bibfnamefont {X.}~\bibnamefont {Qi}}\ and\ \bibinfo {author} {\bibfnamefont {H.}~\bibnamefont {Sun}},\ }\href@noop {} {\  (\bibinfo {year} {2024})},\ \Eprint {http://arxiv.org/abs/2407.21292} {arXiv:2407.21292 [hep-ph]} \BibitemShut {NoStop}%
\bibitem [{\citenamefont {Okada}\ \emph {et~al.}(2012)\citenamefont {Okada}, \citenamefont {Orikasa},\ and\ \citenamefont {Yamada}}]{Okada:2012fs}%
  \BibitemOpen
  \bibfield  {author} {\bibinfo {author} {\bibfnamefont {N.}~\bibnamefont {Okada}}, \bibinfo {author} {\bibfnamefont {Y.}~\bibnamefont {Orikasa}}, \ and\ \bibinfo {author} {\bibfnamefont {T.}~\bibnamefont {Yamada}},\ }\href {\doibase 10.1103/PhysRevD.86.076003} {\bibfield  {journal} {\bibinfo  {journal} {Phys. Rev. D}\ }\textbf {\bibinfo {volume} {86}},\ \bibinfo {pages} {076003} (\bibinfo {year} {2012})},\ \Eprint {http://arxiv.org/abs/1207.1510} {arXiv:1207.1510 [hep-ph]} \BibitemShut {NoStop}%
\bibitem [{\citenamefont {Garny}(2007)}]{Garny:2006xn}%
  \BibitemOpen
  \bibfield  {author} {\bibinfo {author} {\bibfnamefont {M.}~\bibnamefont {Garny}},\ }\href {\doibase 10.1088/1751-8113/40/25/S53} {\bibfield  {journal} {\bibinfo  {journal} {J. Phys. A}\ }\textbf {\bibinfo {volume} {40}},\ \bibinfo {pages} {7005} (\bibinfo {year} {2007})},\ \Eprint {http://arxiv.org/abs/hep-ph/0612145} {arXiv:hep-ph/0612145} \BibitemShut {NoStop}%
\bibitem [{\citenamefont {Morrissey}\ and\ \citenamefont {Ramsey-Musolf}(2012)}]{Morrissey:2012db}%
  \BibitemOpen
  \bibfield  {author} {\bibinfo {author} {\bibfnamefont {D.~E.}\ \bibnamefont {Morrissey}}\ and\ \bibinfo {author} {\bibfnamefont {M.~J.}\ \bibnamefont {Ramsey-Musolf}},\ }\href {\doibase 10.1088/1367-2630/14/12/125003} {\bibfield  {journal} {\bibinfo  {journal} {New J. Phys.}\ }\textbf {\bibinfo {volume} {14}},\ \bibinfo {pages} {125003} (\bibinfo {year} {2012})},\ \Eprint {http://arxiv.org/abs/1206.2942} {arXiv:1206.2942 [hep-ph]} \BibitemShut {NoStop}%
\bibitem [{\citenamefont {Buchmuller}\ \emph {et~al.}(2003)\citenamefont {Buchmuller}, \citenamefont {Di~Bari},\ and\ \citenamefont {Plumacher}}]{Buchmuller:2003gz}%
  \BibitemOpen
  \bibfield  {author} {\bibinfo {author} {\bibfnamefont {W.}~\bibnamefont {Buchmuller}}, \bibinfo {author} {\bibfnamefont {P.}~\bibnamefont {Di~Bari}}, \ and\ \bibinfo {author} {\bibfnamefont {M.}~\bibnamefont {Plumacher}},\ }\href {\doibase 10.1016/S0550-3213(03)00449-8} {\bibfield  {journal} {\bibinfo  {journal} {Nucl. Phys. B}\ }\textbf {\bibinfo {volume} {665}},\ \bibinfo {pages} {445} (\bibinfo {year} {2003})},\ \Eprint {http://arxiv.org/abs/hep-ph/0302092} {arXiv:hep-ph/0302092} \BibitemShut {NoStop}%
\bibitem [{\citenamefont {Dolgov}\ and\ \citenamefont {Freese}(1995)}]{Dolgov:1994zq}%
  \BibitemOpen
  \bibfield  {author} {\bibinfo {author} {\bibfnamefont {A.}~\bibnamefont {Dolgov}}\ and\ \bibinfo {author} {\bibfnamefont {K.}~\bibnamefont {Freese}},\ }\href {\doibase 10.1103/PhysRevD.51.2693} {\bibfield  {journal} {\bibinfo  {journal} {Phys. Rev. D}\ }\textbf {\bibinfo {volume} {51}},\ \bibinfo {pages} {2693} (\bibinfo {year} {1995})},\ \Eprint {http://arxiv.org/abs/hep-ph/9410346} {arXiv:hep-ph/9410346} \BibitemShut {NoStop}%
\bibitem [{\citenamefont {Dolgov}\ \emph {et~al.}(1997)\citenamefont {Dolgov}, \citenamefont {Freese}, \citenamefont {Rangarajan},\ and\ \citenamefont {Srednicki}}]{Dolgov:1996qq}%
  \BibitemOpen
  \bibfield  {author} {\bibinfo {author} {\bibfnamefont {A.}~\bibnamefont {Dolgov}}, \bibinfo {author} {\bibfnamefont {K.}~\bibnamefont {Freese}}, \bibinfo {author} {\bibfnamefont {R.}~\bibnamefont {Rangarajan}}, \ and\ \bibinfo {author} {\bibfnamefont {M.}~\bibnamefont {Srednicki}},\ }\href {\doibase 10.1103/PhysRevD.56.6155} {\bibfield  {journal} {\bibinfo  {journal} {Phys. Rev. D}\ }\textbf {\bibinfo {volume} {56}},\ \bibinfo {pages} {6155} (\bibinfo {year} {1997})},\ \Eprint {http://arxiv.org/abs/hep-ph/9610405} {arXiv:hep-ph/9610405} \BibitemShut {NoStop}%
\bibitem [{\citenamefont {Luongo}\ \emph {et~al.}(2023)\citenamefont {Luongo}, \citenamefont {Marcantognini},\ and\ \citenamefont {Muccino}}]{Luongo:2021gho}%
  \BibitemOpen
  \bibfield  {author} {\bibinfo {author} {\bibfnamefont {O.}~\bibnamefont {Luongo}}, \bibinfo {author} {\bibfnamefont {N.}~\bibnamefont {Marcantognini}}, \ and\ \bibinfo {author} {\bibfnamefont {M.}~\bibnamefont {Muccino}},\ }\href {\doibase 10.1007/s10714-023-03079-7} {\bibfield  {journal} {\bibinfo  {journal} {Gen. Rel. Grav.}\ }\textbf {\bibinfo {volume} {55}},\ \bibinfo {pages} {33} (\bibinfo {year} {2023})},\ \Eprint {http://arxiv.org/abs/2112.05730} {arXiv:2112.05730 [hep-ph]} \BibitemShut {NoStop}%
\bibitem [{\citenamefont {Davoudiasl}\ \emph {et~al.}(2004)\citenamefont {Davoudiasl}, \citenamefont {Kitano}, \citenamefont {Kribs}, \citenamefont {Murayama},\ and\ \citenamefont {Steinhardt}}]{Davoudiasl:2004gf}%
  \BibitemOpen
  \bibfield  {author} {\bibinfo {author} {\bibfnamefont {H.}~\bibnamefont {Davoudiasl}}, \bibinfo {author} {\bibfnamefont {R.}~\bibnamefont {Kitano}}, \bibinfo {author} {\bibfnamefont {G.~D.}\ \bibnamefont {Kribs}}, \bibinfo {author} {\bibfnamefont {H.}~\bibnamefont {Murayama}}, \ and\ \bibinfo {author} {\bibfnamefont {P.~J.}\ \bibnamefont {Steinhardt}},\ }\href {\doibase 10.1103/PhysRevLett.93.201301} {\bibfield  {journal} {\bibinfo  {journal} {Phys. Rev. Lett.}\ }\textbf {\bibinfo {volume} {93}},\ \bibinfo {pages} {201301} (\bibinfo {year} {2004})},\ \Eprint {http://arxiv.org/abs/hep-ph/0403019} {arXiv:hep-ph/0403019} \BibitemShut {NoStop}%
\bibitem [{\citenamefont {Arbuzova}\ \emph {et~al.}(2023)\citenamefont {Arbuzova}, \citenamefont {Dolgov}, \citenamefont {Dutta},\ and\ \citenamefont {Rangarajan}}]{Arbuzova:2023rri}%
  \BibitemOpen
  \bibfield  {author} {\bibinfo {author} {\bibfnamefont {E.}~\bibnamefont {Arbuzova}}, \bibinfo {author} {\bibfnamefont {A.}~\bibnamefont {Dolgov}}, \bibinfo {author} {\bibfnamefont {K.}~\bibnamefont {Dutta}}, \ and\ \bibinfo {author} {\bibfnamefont {R.}~\bibnamefont {Rangarajan}},\ }\href {\doibase 10.3390/sym15020404} {\bibfield  {journal} {\bibinfo  {journal} {Symmetry}\ }\textbf {\bibinfo {volume} {15}},\ \bibinfo {pages} {404} (\bibinfo {year} {2023})},\ \Eprint {http://arxiv.org/abs/2301.08322} {arXiv:2301.08322 [gr-qc]} \BibitemShut {NoStop}%
\bibitem [{\citenamefont {Sadjadi}(2007)}]{Sadjadi:2007dx}%
  \BibitemOpen
  \bibfield  {author} {\bibinfo {author} {\bibfnamefont {H.~M.}\ \bibnamefont {Sadjadi}},\ }\href {\doibase 10.1103/PhysRevD.76.123507} {\bibfield  {journal} {\bibinfo  {journal} {Phys. Rev. D}\ }\textbf {\bibinfo {volume} {76}},\ \bibinfo {pages} {123507} (\bibinfo {year} {2007})},\ \Eprint {http://arxiv.org/abs/0709.0697} {arXiv:0709.0697 [gr-qc]} \BibitemShut {NoStop}%
\bibitem [{\citenamefont {Balazs}(2014)}]{Balazs:2014eba}%
  \BibitemOpen
  \bibfield  {author} {\bibinfo {author} {\bibfnamefont {C.}~\bibnamefont {Balazs}},\ }\href@noop {} {\  (\bibinfo {year} {2014})},\ \Eprint {http://arxiv.org/abs/1411.3398} {arXiv:1411.3398 [hep-ph]} \BibitemShut {NoStop}%
\bibitem [{\citenamefont {Racker}(2016)}]{Racker:2014yfa}%
  \BibitemOpen
  \bibfield  {author} {\bibinfo {author} {\bibfnamefont {J.}~\bibnamefont {Racker}},\ }\href {\doibase 10.1016/j.nuclphysbps.2015.09.047} {\bibfield  {journal} {\bibinfo  {journal} {Nucl. Part. Phys. Proc.}\ }\textbf {\bibinfo {volume} {273-275}},\ \bibinfo {pages} {334} (\bibinfo {year} {2016})},\ \Eprint {http://arxiv.org/abs/1410.5482} {arXiv:1410.5482 [hep-ph]} \BibitemShut {NoStop}%
\bibitem [{\citenamefont {Goodarzi}(2023)}]{Goodarzi:2023ltp}%
  \BibitemOpen
  \bibfield  {author} {\bibinfo {author} {\bibfnamefont {P.}~\bibnamefont {Goodarzi}},\ }\href {\doibase 10.1140/epjc/s10052-023-12182-7} {\bibfield  {journal} {\bibinfo  {journal} {Eur. Phys. J. C}\ }\textbf {\bibinfo {volume} {83}},\ \bibinfo {pages} {990} (\bibinfo {year} {2023})},\ \Eprint {http://arxiv.org/abs/2307.10709} {arXiv:2307.10709 [hep-th]} \BibitemShut {NoStop}%
\bibitem [{\citenamefont {Arbuzova}\ and\ \citenamefont {Dolgov}(2019)}]{Arbuzova:2017zby}%
  \BibitemOpen
  \bibfield  {author} {\bibinfo {author} {\bibfnamefont {E.~V.}\ \bibnamefont {Arbuzova}}\ and\ \bibinfo {author} {\bibfnamefont {A.~D.}\ \bibnamefont {Dolgov}},\ }in\ \href {\doibase 10.1142/9789811202339_0059} {\emph {\bibinfo {booktitle} {{18th Lomonosov Conference on Elementary Particle Physics}}}}\ (\bibinfo {year} {2019})\ pp.\ \bibinfo {pages} {309--313},\ \Eprint {http://arxiv.org/abs/1712.04627} {arXiv:1712.04627 [hep-ph]} \BibitemShut {NoStop}%
\bibitem [{\citenamefont {Mojahed}\ \emph {et~al.}(2025)\citenamefont {Mojahed}, \citenamefont {Schmitz},\ and\ \citenamefont {Xu}}]{Mojahed:2024mvb}%
  \BibitemOpen
  \bibfield  {author} {\bibinfo {author} {\bibfnamefont {M.~A.}\ \bibnamefont {Mojahed}}, \bibinfo {author} {\bibfnamefont {K.}~\bibnamefont {Schmitz}}, \ and\ \bibinfo {author} {\bibfnamefont {X.-J.}\ \bibnamefont {Xu}},\ }\href {\doibase 10.1103/PhysRevD.111.055005} {\bibfield  {journal} {\bibinfo  {journal} {Phys. Rev. D}\ }\textbf {\bibinfo {volume} {111}},\ \bibinfo {pages} {055005} (\bibinfo {year} {2025})},\ \Eprint {http://arxiv.org/abs/2409.10605} {arXiv:2409.10605 [hep-ph]} \BibitemShut {NoStop}%
\bibitem [{\citenamefont {Akrami}\ \emph {et~al.}(2020)\citenamefont {Akrami} \emph {et~al.}}]{Planck:2018jri}%
  \BibitemOpen
  \bibfield  {author} {\bibinfo {author} {\bibfnamefont {Y.}~\bibnamefont {Akrami}} \emph {et~al.} (\bibinfo {collaboration} {Planck}),\ }\href {\doibase 10.1051/0004-6361/201833887} {\bibfield  {journal} {\bibinfo  {journal} {Astron. Astrophys.}\ }\textbf {\bibinfo {volume} {641}},\ \bibinfo {pages} {A10} (\bibinfo {year} {2020})},\ \Eprint {http://arxiv.org/abs/1807.06211} {arXiv:1807.06211 [astro-ph.CO]} \BibitemShut {NoStop}%
\bibitem [{\citenamefont {Belfiglio}\ \emph {et~al.}(2023{\natexlab{a}})\citenamefont {Belfiglio}, \citenamefont {Giamb\`o},\ and\ \citenamefont {Luongo}}]{Belfiglio:2022qai}%
  \BibitemOpen
  \bibfield  {author} {\bibinfo {author} {\bibfnamefont {A.}~\bibnamefont {Belfiglio}}, \bibinfo {author} {\bibfnamefont {R.}~\bibnamefont {Giamb\`o}}, \ and\ \bibinfo {author} {\bibfnamefont {O.}~\bibnamefont {Luongo}},\ }\href {\doibase 10.1088/1361-6382/accc00} {\bibfield  {journal} {\bibinfo  {journal} {Class. Quant. Grav.}\ }\textbf {\bibinfo {volume} {40}},\ \bibinfo {pages} {105004} (\bibinfo {year} {2023}{\natexlab{a}})},\ \Eprint {http://arxiv.org/abs/2206.14158} {arXiv:2206.14158 [gr-qc]} \BibitemShut {NoStop}%
\bibitem [{\citenamefont {Belfiglio}\ \emph {et~al.}(2024{\natexlab{a}})\citenamefont {Belfiglio}, \citenamefont {Carloni},\ and\ \citenamefont {Luongo}}]{Belfiglio:2023rxb}%
  \BibitemOpen
  \bibfield  {author} {\bibinfo {author} {\bibfnamefont {A.}~\bibnamefont {Belfiglio}}, \bibinfo {author} {\bibfnamefont {Y.}~\bibnamefont {Carloni}}, \ and\ \bibinfo {author} {\bibfnamefont {O.}~\bibnamefont {Luongo}},\ }\href {\doibase 10.1016/j.dark.2024.101458} {\bibfield  {journal} {\bibinfo  {journal} {Phys. Dark Univ.}\ }\textbf {\bibinfo {volume} {44}},\ \bibinfo {pages} {101458} (\bibinfo {year} {2024}{\natexlab{a}})},\ \Eprint {http://arxiv.org/abs/2307.04739} {arXiv:2307.04739 [gr-qc]} \BibitemShut {NoStop}%
\bibitem [{\citenamefont {Luongo}\ and\ \citenamefont {Mengoni}(2024)}]{Luongo:2024opv}%
  \BibitemOpen
  \bibfield  {author} {\bibinfo {author} {\bibfnamefont {O.}~\bibnamefont {Luongo}}\ and\ \bibinfo {author} {\bibfnamefont {T.}~\bibnamefont {Mengoni}},\ }\href {\doibase 10.1088/1361-6382/ad3ac9} {\bibfield  {journal} {\bibinfo  {journal} {Class. Quant. Grav.}\ }\textbf {\bibinfo {volume} {41}},\ \bibinfo {pages} {105006} (\bibinfo {year} {2024})}\BibitemShut {NoStop}%
\bibitem [{\citenamefont {Belfiglio}\ \emph {et~al.}(2024{\natexlab{b}})\citenamefont {Belfiglio}, \citenamefont {Luongo},\ and\ \citenamefont {Mengoni}}]{Belfiglio:2024swy}%
  \BibitemOpen
  \bibfield  {author} {\bibinfo {author} {\bibfnamefont {A.}~\bibnamefont {Belfiglio}}, \bibinfo {author} {\bibfnamefont {O.}~\bibnamefont {Luongo}}, \ and\ \bibinfo {author} {\bibfnamefont {T.}~\bibnamefont {Mengoni}},\ }\href@noop {} {\  (\bibinfo {year} {2024}{\natexlab{b}})},\ \Eprint {http://arxiv.org/abs/2411.11130} {arXiv:2411.11130 [gr-qc]} \BibitemShut {NoStop}%
\bibitem [{\citenamefont {Luongo}\ and\ \citenamefont {Muccino}(2018)}]{Luongo:2018lgy}%
  \BibitemOpen
  \bibfield  {author} {\bibinfo {author} {\bibfnamefont {O.}~\bibnamefont {Luongo}}\ and\ \bibinfo {author} {\bibfnamefont {M.}~\bibnamefont {Muccino}},\ }\href {\doibase 10.1103/PhysRevD.98.103520} {\bibfield  {journal} {\bibinfo  {journal} {Phys. Rev. D}\ }\textbf {\bibinfo {volume} {98}},\ \bibinfo {pages} {103520} (\bibinfo {year} {2018})},\ \Eprint {http://arxiv.org/abs/1807.00180} {arXiv:1807.00180 [gr-qc]} \BibitemShut {NoStop}%
\bibitem [{\citenamefont {D'Agostino}\ \emph {et~al.}(2022)\citenamefont {D'Agostino}, \citenamefont {Luongo},\ and\ \citenamefont {Muccino}}]{DAgostino:2022fcx}%
  \BibitemOpen
  \bibfield  {author} {\bibinfo {author} {\bibfnamefont {R.}~\bibnamefont {D'Agostino}}, \bibinfo {author} {\bibfnamefont {O.}~\bibnamefont {Luongo}}, \ and\ \bibinfo {author} {\bibfnamefont {M.}~\bibnamefont {Muccino}},\ }\href {\doibase 10.1088/1361-6382/ac8af2} {\bibfield  {journal} {\bibinfo  {journal} {Class. Quant. Grav.}\ }\textbf {\bibinfo {volume} {39}},\ \bibinfo {pages} {195014} (\bibinfo {year} {2022})},\ \Eprint {http://arxiv.org/abs/2204.02190} {arXiv:2204.02190 [gr-qc]} \BibitemShut {NoStop}%
\bibitem [{\citenamefont {Sakharov}(1967)}]{Sakharov:1967dj}%
  \BibitemOpen
  \bibfield  {author} {\bibinfo {author} {\bibfnamefont {A.~D.}\ \bibnamefont {Sakharov}},\ }\href {\doibase 10.1070/PU1991v034n05ABEH002497} {\bibfield  {journal} {\bibinfo  {journal} {Pisma Zh. Eksp. Teor. Fiz.}\ }\textbf {\bibinfo {volume} {5}},\ \bibinfo {pages} {32} (\bibinfo {year} {1967})}\BibitemShut {NoStop}%
\bibitem [{\citenamefont {Dimopoulos}\ and\ \citenamefont {Susskind}(1978)}]{Dimopoulos:1978kv}%
  \BibitemOpen
  \bibfield  {author} {\bibinfo {author} {\bibfnamefont {S.}~\bibnamefont {Dimopoulos}}\ and\ \bibinfo {author} {\bibfnamefont {L.}~\bibnamefont {Susskind}},\ }\href {\doibase 10.1103/PhysRevD.18.4500} {\bibfield  {journal} {\bibinfo  {journal} {Phys. Rev. D}\ }\textbf {\bibinfo {volume} {18}},\ \bibinfo {pages} {4500} (\bibinfo {year} {1978})}\BibitemShut {NoStop}%
\bibitem [{\citenamefont {Weinberg}(1979)}]{Weinberg:1979bt}%
  \BibitemOpen
  \bibfield  {author} {\bibinfo {author} {\bibfnamefont {S.}~\bibnamefont {Weinberg}},\ }\href {\doibase 10.1103/PhysRevLett.42.850} {\bibfield  {journal} {\bibinfo  {journal} {Phys. Rev. Lett.}\ }\textbf {\bibinfo {volume} {42}},\ \bibinfo {pages} {850} (\bibinfo {year} {1979})}\BibitemShut {NoStop}%
\bibitem [{\citenamefont {De~Simone}\ and\ \citenamefont {Kobayashi}(2016)}]{DeSimone:2016ofp}%
  \BibitemOpen
  \bibfield  {author} {\bibinfo {author} {\bibfnamefont {A.}~\bibnamefont {De~Simone}}\ and\ \bibinfo {author} {\bibfnamefont {T.}~\bibnamefont {Kobayashi}},\ }\href {\doibase 10.1088/1475-7516/2016/08/052} {\bibfield  {journal} {\bibinfo  {journal} {JCAP}\ }\textbf {\bibinfo {volume} {08}},\ \bibinfo {pages} {052} (\bibinfo {year} {2016})},\ \Eprint {http://arxiv.org/abs/1605.00670} {arXiv:1605.00670 [hep-ph]} \BibitemShut {NoStop}%
\bibitem [{\citenamefont {Stein}\ and\ \citenamefont {Kinney}(2022)}]{Stein:2021uge}%
  \BibitemOpen
  \bibfield  {author} {\bibinfo {author} {\bibfnamefont {N.~K.}\ \bibnamefont {Stein}}\ and\ \bibinfo {author} {\bibfnamefont {W.~H.}\ \bibnamefont {Kinney}},\ }\href {\doibase 10.1088/1475-7516/2022/01/022} {\bibfield  {journal} {\bibinfo  {journal} {JCAP}\ }\textbf {\bibinfo {volume} {01}},\ \bibinfo {pages} {022} (\bibinfo {year} {2022})},\ \Eprint {http://arxiv.org/abs/2106.02089} {arXiv:2106.02089 [astro-ph.CO]} \BibitemShut {NoStop}%
\bibitem [{\citenamefont {Adams}\ \emph {et~al.}(1993)\citenamefont {Adams}, \citenamefont {Bond}, \citenamefont {Freese}, \citenamefont {Frieman},\ and\ \citenamefont {Olinto}}]{Adams:1992bn}%
  \BibitemOpen
  \bibfield  {author} {\bibinfo {author} {\bibfnamefont {F.~C.}\ \bibnamefont {Adams}}, \bibinfo {author} {\bibfnamefont {J.~R.}\ \bibnamefont {Bond}}, \bibinfo {author} {\bibfnamefont {K.}~\bibnamefont {Freese}}, \bibinfo {author} {\bibfnamefont {J.~A.}\ \bibnamefont {Frieman}}, \ and\ \bibinfo {author} {\bibfnamefont {A.~V.}\ \bibnamefont {Olinto}},\ }\href {\doibase 10.1103/PhysRevD.47.426} {\bibfield  {journal} {\bibinfo  {journal} {Phys. Rev. D}\ }\textbf {\bibinfo {volume} {47}},\ \bibinfo {pages} {426} (\bibinfo {year} {1993})},\ \Eprint {http://arxiv.org/abs/hep-ph/9207245} {arXiv:hep-ph/9207245} \BibitemShut {NoStop}%
\bibitem [{\citenamefont {Opferkuch}\ \emph {et~al.}(2019)\citenamefont {Opferkuch}, \citenamefont {Schwaller},\ and\ \citenamefont {Stefanek}}]{Opferkuch:2019zbd}%
  \BibitemOpen
  \bibfield  {author} {\bibinfo {author} {\bibfnamefont {T.}~\bibnamefont {Opferkuch}}, \bibinfo {author} {\bibfnamefont {P.}~\bibnamefont {Schwaller}}, \ and\ \bibinfo {author} {\bibfnamefont {B.~A.}\ \bibnamefont {Stefanek}},\ }\href {\doibase 10.1088/1475-7516/2019/07/016} {\bibfield  {journal} {\bibinfo  {journal} {JCAP}\ }\textbf {\bibinfo {volume} {07}},\ \bibinfo {pages} {016} (\bibinfo {year} {2019})},\ \Eprint {http://arxiv.org/abs/1905.06823} {arXiv:1905.06823 [gr-qc]} \BibitemShut {NoStop}%
\bibitem [{\citenamefont {Belfiglio}\ and\ \citenamefont {Luongo}(2025)}]{Belfiglio:2025chv}%
  \BibitemOpen
  \bibfield  {author} {\bibinfo {author} {\bibfnamefont {A.}~\bibnamefont {Belfiglio}}\ and\ \bibinfo {author} {\bibfnamefont {O.}~\bibnamefont {Luongo}},\ }\href@noop {} {\  (\bibinfo {year} {2025})},\ \Eprint {http://arxiv.org/abs/2504.04219} {arXiv:2504.04219 [gr-qc]} \BibitemShut {NoStop}%
\bibitem [{\citenamefont {Belfiglio}\ and\ \citenamefont {Luongo}(2024)}]{Belfiglio:2024xqt}%
  \BibitemOpen
  \bibfield  {author} {\bibinfo {author} {\bibfnamefont {A.}~\bibnamefont {Belfiglio}}\ and\ \bibinfo {author} {\bibfnamefont {O.}~\bibnamefont {Luongo}},\ }\href {\doibase 10.1103/PhysRevD.110.023541} {\bibfield  {journal} {\bibinfo  {journal} {Phys. Rev. D}\ }\textbf {\bibinfo {volume} {110}},\ \bibinfo {pages} {023541} (\bibinfo {year} {2024})},\ \Eprint {http://arxiv.org/abs/2401.16910} {arXiv:2401.16910 [hep-th]} \BibitemShut {NoStop}%
\bibitem [{\citenamefont {Belfiglio}\ \emph {et~al.}(2024{\natexlab{c}})\citenamefont {Belfiglio}, \citenamefont {Luongo},\ and\ \citenamefont {Mancini}}]{Belfiglio:2023moe}%
  \BibitemOpen
  \bibfield  {author} {\bibinfo {author} {\bibfnamefont {A.}~\bibnamefont {Belfiglio}}, \bibinfo {author} {\bibfnamefont {O.}~\bibnamefont {Luongo}}, \ and\ \bibinfo {author} {\bibfnamefont {S.}~\bibnamefont {Mancini}},\ }\href {\doibase 10.1103/PhysRevD.109.123520} {\bibfield  {journal} {\bibinfo  {journal} {Phys. Rev. D}\ }\textbf {\bibinfo {volume} {109}},\ \bibinfo {pages} {123520} (\bibinfo {year} {2024}{\natexlab{c}})},\ \Eprint {http://arxiv.org/abs/2312.11419} {arXiv:2312.11419 [gr-qc]} \BibitemShut {NoStop}%
\bibitem [{\citenamefont {Belfiglio}\ \emph {et~al.}(2023{\natexlab{b}})\citenamefont {Belfiglio}, \citenamefont {Luongo},\ and\ \citenamefont {Mancini}}]{Belfiglio:2022yvs}%
  \BibitemOpen
  \bibfield  {author} {\bibinfo {author} {\bibfnamefont {A.}~\bibnamefont {Belfiglio}}, \bibinfo {author} {\bibfnamefont {O.}~\bibnamefont {Luongo}}, \ and\ \bibinfo {author} {\bibfnamefont {S.}~\bibnamefont {Mancini}},\ }\href {\doibase 10.1103/PhysRevD.107.103512} {\bibfield  {journal} {\bibinfo  {journal} {Phys. Rev. D}\ }\textbf {\bibinfo {volume} {107}},\ \bibinfo {pages} {103512} (\bibinfo {year} {2023}{\natexlab{b}})},\ \Eprint {http://arxiv.org/abs/2212.06448} {arXiv:2212.06448 [gr-qc]} \BibitemShut {NoStop}%
\bibitem [{\citenamefont {Belfiglio}\ \emph {et~al.}(2022)\citenamefont {Belfiglio}, \citenamefont {Luongo},\ and\ \citenamefont {Mancini}}]{Belfiglio:2022cnd}%
  \BibitemOpen
  \bibfield  {author} {\bibinfo {author} {\bibfnamefont {A.}~\bibnamefont {Belfiglio}}, \bibinfo {author} {\bibfnamefont {O.}~\bibnamefont {Luongo}}, \ and\ \bibinfo {author} {\bibfnamefont {S.}~\bibnamefont {Mancini}},\ }\href {\doibase 10.1103/PhysRevD.105.123523} {\bibfield  {journal} {\bibinfo  {journal} {Phys. Rev. D}\ }\textbf {\bibinfo {volume} {105}},\ \bibinfo {pages} {123523} (\bibinfo {year} {2022})},\ \Eprint {http://arxiv.org/abs/2201.12299} {arXiv:2201.12299 [gr-qc]} \BibitemShut {NoStop}%
\bibitem [{\citenamefont {Aviles}\ \emph {et~al.}(2011)\citenamefont {Aviles}, \citenamefont {Bonanno}, \citenamefont {Luongo},\ and\ \citenamefont {Quevedo}}]{Aviles:2011sfa}%
  \BibitemOpen
  \bibfield  {author} {\bibinfo {author} {\bibfnamefont {A.}~\bibnamefont {Aviles}}, \bibinfo {author} {\bibfnamefont {L.}~\bibnamefont {Bonanno}}, \bibinfo {author} {\bibfnamefont {O.}~\bibnamefont {Luongo}}, \ and\ \bibinfo {author} {\bibfnamefont {H.}~\bibnamefont {Quevedo}},\ }\href {\doibase 10.1103/PhysRevD.84.103520} {\bibfield  {journal} {\bibinfo  {journal} {Phys. Rev. D}\ }\textbf {\bibinfo {volume} {84}},\ \bibinfo {pages} {103520} (\bibinfo {year} {2011})},\ \Eprint {http://arxiv.org/abs/1109.3177} {arXiv:1109.3177 [gr-qc]} \BibitemShut {NoStop}%
\end{thebibliography}
\end{document}